\documentclass[preprint, 12pt]{emulateapj}
\usepackage{natbib}
\begin{document}

\title{The 3.3 $\micron$ PAH emission as a star formation rate indicator} 
\author{Ji Hoon Kim$^{1,2}$ Myungshin Im$^{1,3}$, Hyung Mok Lee$^{3}$, Myung Gyoon Lee$^{3}$, Hyunsung David Jun$^{1}$, 
Takao Nakagawa$^{4}$, Hideo Matsuhara$^{4}$, Takehiko Wada$^{4}$, Shinki Oyabu$^{5}$, Toshinobu Takagi$^{4}$, Hanae Inami$^{7}$, 
Youichi Ohyama$^{6}$, Rika Yamada$^{5}$, George Helou$^{8}$, Lee Armus$^{7}$, and Yong Shi$^{8}$}
\affil{$^{1}$Center for the Exploration of the Origin of the Universe,
Astronomy Program, Department of Physics and Astronomy,\\
Seoul National University, Seoul, Republic of Korea\\
$^{2}$National Research Foundation of Korea Postdoctoral Fellow\\
$^{3}$Astronomy Program, Department of Physics and Astronomy, FPRD, Seoul National University, Seoul, Republic of Korea\\
$^{4}$Institute of Space and Astronautical Science, Japan Aerospace Exploration Agency, Sagamihara, Kanagawa 252-5210, Japan\\
$^{5}$Institute for Advanced Research, Nagoya University, Furo-cho, Chikusa-ku, Nagoya 464-8601, Japan\\
$^{6}$Academia Sinica, Institute of Astronomy and Astrophysics, P.O. Box 23-141, Taipei 10617, Taiwan, R.O.C.\\
$^{7}$Spitzer Science Center, California Institute of Technology, Pasadena, CA 91125, USA\\
$^{8}$Infrared Processing and Analysis Center, California Institute of Technology, 1200 East California Boulevard, Pasadena, CA 91125, USA\\
}
\email{jhkim@astro.snu.ac.kr, mim@astro.snu.ac.kr}

\begin{abstract}

Polycyclic Aromatic Hydrocarbon (PAH) emission features dominate the mid-infrared spectra of star-forming galaxies and can be useful to calibrate star formation rates and diagnose ionized states of grains. However, the PAH 3.3 $\mu m$ feature has not been studied as much as other PAH features since it is weaker than others and resides outside of $Spitzer$ capability.
In order to detect and calibrate the 3.3 $\mu m$ PAH emission and investigate its potential as a star formation rate indicator, we carried out an AKARI mission program, AKARI mJy Unbiased Survey of Extragalactic Survey (AMUSES) and compare its sample with various literature samples. We obtained 2 $\sim$ 5 $\micron$ low resolution spectra of 20 flux-limited galaxies with mixed SED classes, which yields the detection of the 3.3 $\mu m$ PAH emission from three out of 20 galaxies. For the combined sample of AMUSES and literature samples, the 3.3 $\micron$ PAH luminosities correlate with the infrared luminosities of star-forming galaxies, albeit with a large scatter (~1.5 dex). The correlation appears to break down at the domain of ultra-luminous infrared galaxies (ULIRGs), and the power of the 3.3 $\micron$ PAH luminosity as a proxy for the infrared luminosity is hampered at log[$L_{\mathrm{PAH3.3}}$/erg/sec$^{-1}$] $>$  $\sim$42.0.
Possible origins for this deviation in the correlation are discussed, including contributions from AGN and strongly obscured YSOs, and the destruction of PAH molecules in ULIRGs.
\end{abstract}

\section{Introduction}
\label{intro}

Over the past decades, infrared (IR) astronomy made a tremendous progress largely thanks to various space missions, such as the Infra-Red Astronomical Satellite (IRAS: \citealt{soifer87}), the Infrared Space Observatory (ISO;  \citealt{Ks96,Gz00}), and the $Spitzer$ space telescope  \citep{spitzer}.  
The most important aspect carried by IR wavelengths is probably that IR emission represents the dust-obscured star formation activity of galaxies \citep{Gz00}.  The bolometric IR luminosity of galaxies measures the dust-obscured star formation within galaxies and is less affected by extinction while other shorter wavelength star formation rate (SFR) proxies falling on optical and ultra-violet (UV) wavelength regimes.
However using the bolometric IR luminosity as a star formation (SF) indicator has two caveats. First, not only newly formed massive stars, but also evolved stellar populations can heat dust components within galaxies \citep{Cz11}. Then it is extremely tricky to obtain the whole range of IR spectral energy distribution (SED) for high-z galaxies to measure bolometric IR luminosities.
Therefore, many studies have attempted calibrate several IR SFR proxy candidates including 8 $\micron$ and 24 $\micron$ bands to bolometric IR luminosities and other SFR indicators in order to establish reliable IR SFR proxies in recent years. 

Polycyclic aromatic hydrocarbons (PAHs) have gotten enormous attention due to their ubiquity and strong potential as diagnostics of other properties such as ionized states and sizes of grains. PAHs are considered to be present in a wide range of objects and environments, such as post-AGB stars, planetary nebulae, HII regions, reflection nebulae and the diffuse interstellar medium  \citep{Pg85,Al89}.
The PAH features are believed to contribute up to 10 \% of the total IR luminosity of star forming galaxies  \citep{He01,Pt02,Sm07}.

Numerous recent studies measure PAH band fluxes and equivalent widths (EWs) in order to calibrate these emission features as SFR proxies within the Galactic environments and galaxies at higher redshift. These studies reveal that there exist differences in PAH EWs and L$_{PAH}$/$L_{\mathrm{IR}}$ ratios between local values and high redshift ones  \citep{He01,Pt02, Sm07}. Since PAH band ratios reflect variations in physical conditions within environments, such as ionization states of dust grains and metallicity  \citep{Sm07,Ga08,Go08, Dr11,Ga11}, more detailed study on this subject will put a better constraint on physical conditions of PAH emission sites and calibration of PAH bands as SFR proxies.

These previous studies on PAH emission features concentrate on stronger bands, such as 6.2, 7.7, and 11.3 $\mu m$. On the other hand, studies on 3.3 $\mu m$ feature have been far fewer, due to its relatively weaker strength. There still have been efforts to investigate the 3.3 $\micron$ PAH emission and its property related to star formation activity. The first of such studies is \citet{To91}. Compiling infrared L-band (3-4 $\micron$) spectra from various sources, \citet{To91} categorized the 3.3 $\micron$ PAH emission feature into two types and investigated if the origin of the emission feature is PAHs. Several studies on energy source of IR emission followed. Analyzing 57 AGN and one starburst galaxy spectra, \citet{Cl00} detected the 3.3 $\micron$ PAH emission from 47 of them and used its strength for classifying the target spectra. 
\citet{RAV03} detected the 3.3 $\micron$ emission in two Seyfert 1 galaxies and one quasi-stellar object (QSO) using $NASA$ infrared telescope facility (IRTF) SpeX and claimed that these active galactic nuclei (AGN) have the 3.3 $\micron$ PAH luminosity levels similar to those of starburst and LIRGs. \citet{Im03} and \citet{Im04} observed 32 Seyfert 2 galaxies and 23 Seyfert 1 galaxies, respectively,  in order to investigate connection between nuclear starburst activity and AGN activity using ground-based L-band spectroscopy. They detected the 3.3 $\micron$ PAH emission from 10 out of 23 Seyfert 1 galaxies and 11 out of 32 Seyfert 2 galaxies respectively, and found that starburst activity correlates with nuclear activity regardless of types of Seyfert galaxies. Using ISO/SWS spectra of a wide variety of sources, \citet{vD04} claimed that the 3.3 $\micron$ emission originate mainly in neutral and/or negatively charged PAHs in contrast to the 6.2 and 7.7 $\micron$ emission which are from PAH cations.
On the other hand, \citet{WKI08} and \citet{Oi10} confirmed that there is a strong correlation between nuclear starburst activity and AGN activity traced by either X-ray luminosity, or nuclear N-band luminosity for Seyfert galaxies, while detecting the 3.3 $\micron$ PAH emission from three out of eight sources and five out of 22 sources, respectively. Extending these works into more luminous regime, \citet{Im11} observed 30 PG QSOs and detected the 3.3 $\micron$ PAH emission from five QSOs utilizing slit-less spectroscopic capability of AKARI infrared satellite \citep{AKARI} to probe global star formation activity by the 3.3 $\micron$ PAH emission feature.. They confirmed that the correlation between nuclear starburst activity and AGN activity are intact for PG QSOs.

There have been other studies utilizing slit-less spectroscopic capability of AKARI. \citet{Im08} obtained AKARI Infrared Camera (IRC; \citealt{IRC}) spectra of 45 nearby ULIRGs in order to investigate the energy source of ULIRGs. They detected the 3.3 $\micron$ PAH emission from 40 ULIRGs. However, they claimed that the obscured starburst activity is not the dominant energy source for these ULIRGs, even for ULIRGs which are classified as non-Seyfert optically. \citet{Im10} also obtained IRC spectroscopy of 64 LIRGs and 54 ULIRGs and detected the 3.3 $\micron$ PAH emission from the majority of the observed targets. They found that the 3.3 $\micron$ PAH gives a good estimate for SFR as Br$\alpha$ ($\lambda_{rest}$ = 4.05 $\micron$) does. For more of higher redshift objects, \citet{Sj09} detected the 3.3 $\micron$ PAH emission for four out of 11 z $\sim$ 2 ULIRGs.

Understanding how the 3.3 $\mu m$ PAH emission is related to star formation activities has a great importance, since the 3.3 $\mu m$ PAH feature is likely to be the only dust emission feature at high redshift ($z > 4.5$) easily accessible with future space IR telescope missions, such as James Webb Space Telescope (JWST) and SPICA.
Given the importance of understanding 2.5-5 $\mu m$ mid-infrared (MIR) emissions from extragalactic sources, we carried out a study of low redshift objects in the 2.5-5 $\mu m$ window, AKARI mJy Unbiased Survey of Extragalactic Sources (AMUSES) as one of AKARI mission projects (MPs). The main scientific goal of AMUSES is to construct a continuous spectral library over the wavelength window between 2.5 and 40 $\mu m$ for a subsample of 5$mJy$ Unbiased $Spitzer$ Extragalactic Survey (5MUSES,  \citealt{5muses}) by combining spectra from AKARI and $Spitzer$.

In \S \ref{data}, we present sample selection, data acquisition and reduction. Then we present and discuss fitting methodology as well as stacking analysis in \S \ref{analysis}. In \S \ref{results}, we present both reduced individual spectra and the stacked spectra of the sample. We also present and discuss fitting results in  \S \ref{results}. Implication of our results are presented in \S \ref{discussion}.  We summarize our study in \S \ref{sum}.

Throughout this paper, we assume that the universe is flat with $\Omega_{M}$= 0.3 and $\Omega_{\Lambda}$=0.7, and $H_{0}$=70km s$^{-1}$ Mpc$^{-1}$ (e.g., \citealt{Im97}).



\section{Data}
\label{data}

\subsection{Sample Selection}
\label{sample}
The AMUSES sample is drawn from 5MUSES. We choose 5MUSES as our parent sample, because it contains a statistically unbiased library of MIR spectra of IR sources that covers the gap between the bright, nearby IR galaxies and the much fainter, more distant IR sources being studied by the $Spizter/AKARI$.
5MUSES, one of the $Spitzer$ Legacy surveys, performed a MIR spectroscopic observation of extragalactic sources brighter than 5$mJy$ at 24$\mu m$ in the $Spitzer$ First Look Survey (FLS; \citealt{fls, fadafls, yanfls, apfls, choifls, shim1, shim2}) field and four subfields of $Spitzer$ Wide-area Infrared Extragalactic survey (SWIRE; \citealt{swire}) with the Infrared Spectrograph (IRS) onboard the $Spitzer$ space telescope. The main scientific goal of 5MUSES is to provide an unbiased library of infrared spectra from 5 to 40 $\mu m$ of sources which have not been sought after in previous studies.

Since the main objective of AMUSES is to detect the 3.3 $\mu m$ PAH feature, we select 5MUSES galaxies that are brighter than 1 mJy at 3.6 $\mu m$. Based on the $Spitzer$ IRAC 3.6 $\mu m$ data for the 5MUSES sample, we find that 60 of the 330 5MUSES galaxies satisfy the flux cut of 1mJy at 3.6 $\mu m$. We limit the redshift range to z $<$ 0.5 in order to achieve S/N $>$ 5 for detection of the 3.3 $\mu m$ emission feature. With this additional cut, the sample size is reduced to 50 galaxies. In addition to this base sample, we add 10 5MUSES galaxies  with their 3.6 $\mu m$ flux brighter than 0.7 mJy whose redshifts could not be determined by optical spectroscopy. These are added as secondary targets in case that the scheduling constraint makes it impossible to observe the main targets. With these additional galaxies, the original sample consists of  60 galaxies.

Ultimately, a total of 44 galaxies among these 60 galaxies are approved for the final program, after resolving visibility conflicts with targets from other approved programs. However, only 20 target galaxies were observed until the mechanical failure of the AKARI cooling system. The termination of the AKARI scientific mission is announced on 2011 June 20, thus, these 20 galaxies are the final sample for AMUSES. For our final sample, we adopted their SED classes from \citet{5muses}. They classified objects using the 6.2 $\micron$ PAH equivalent widths (EWs): sources with EW $>$ 0.5 $\micron$ are starburst (SB)-dominated; sources with 0.2 $<$ EW $<$ 0.5 $\micron$ are AGN-SB composite and sources with EWs $\le$ 0.2 $\micron$ are AGN-dominated. There are 12 AGNs and seven starburst galaxies within the final sample. One galaxy has a composite SED. The final sample also includes two AGNs at z $>$ 0.7 whose redshifts were determined after our sample selection by \citet{5muses}. The basic properties of the sample are listed in Table \ref{table1}, while the redshift versus the 3.6 $\mu m$ flux is presented in Figure \ref{fig1}.

\subsection{Observation}
\label{obs}

The observation was carried out with the spectroscopy mode of IRC on AKARI. The NIR grism (NG) mode was adopted for most of the observations providing the spectral resolution of R $\sim$ 120 \citep{IRC}. For eight of the targets, NIR prism (NP) mode was used which provide the spectral resolution of R $\sim$ 20. The observations were carried out using 1$\arcmin \times 1\arcmin$  slit aperture to avoid confusion with surrounding sources. The pixel scale of NIR camera is  1.45$\arcsec$ and the full width half maxima of the point spread function is 3.2 pixels which corresponds to 4.64$\arcsec$.
All target galaxies have sizes small enough to fit in the slit aperture, although size variation is not insignificant. We present the R-band images of the target galaxies along with the 1D reduced spectra and the 2D spectrum images in Figure \ref{fig2}. For several galaxies without ancillary R-band images, $Spitzer$ IRAC 3.6 $\micron$ band images are presented. We discuss the extraction width and the target galaxy sizes in section \ref{dr}. 
The actual exposure times run between six to seven minutes  for each pointing observation. Originally, three pointing observations were planned for all the targets, but a half of the targets presented in Table 1 had less than three pointing observations due to the termination of the AKARI science mission. In total, observation of 51 pointings was carried out for 20 target galaxies. Table \ref{table1} also lists how many pointings are acquired for each target.

\subsection{Data Reduction}
\label{dr}

Data reduction was performed with the IRC spectroscopy pipeline.\footnotemark[3]  The IRC spectroscopy pipeline subtracts scaled super-dark frames, applies linearity correction while masking saturation and monochromatic flat-fielding, subtracts background from individual sub-frames, registers and stacks images before subtracting background again from stacked images and extracting 2D spectra. Then wavelength calibration, flat color-term correction, spectral tilt correction, and spectral response calibration were applied on these extracted 2D spectra.  Additional cosmic ray removal, stacking multi-pointing exposures, and one-sigma clipping during stacking were executed individually after processing the data through the IRC spectroscopy pipeline. 

Final one-dimensional (1D) spectra were extracted from two-dimensional (2D) spectral images which are binned by three pixels along the wavelength direction. The 1D spectra were extracted from the 2D spectrum images with various widths in spatial direction. First, we extracted the 1D spectra with the width of 5 pixels which corresponds to 7.3$\arcsec$. In physical scales, this ranges from 1.8 kpc for the nearest target galaxy to 56.3 kpc for the farthest target galaxy. We present the extraction boxes with 5 pixel width over the R-, or IRAC 3.6 $\micron$ band images in Figure \ref{fig2}. As indicated in these figures, these extraction regions cover entire galaxies in most cases. We tried to change the extraction width in order to check if a larger extraction width unveils any hidden PAH 3.3 $\micron$ emission feature from them,  or it changes the extracted PHA flux more than the measurement errors. Specifically, we also applied extraction boxes with 10 pixels and 15 pixels. However, we did not 
notice any significant change of the extracted spectra over the range of extraction width for our entire sample. 
Therefore, we use only the spectra extracted with 5 pixel extraction width for our analysis.

For more details regarding data reduction process, please refer \citet{IRCpipe}.

\footnotetext[3]{http://www.ir.isas.jaxa.jp/ASTRO-F/Observation/DataReduction/IRC/}

\section{Analysis}
\label{analysis}

\subsection{Spectra of the sample and their analysis}
\label{spectra}

We present the reduced spectra and fitting of emission features of 20 AMUSES galaxies in Figure \ref{fig2}. For FBQS J0216-0444 and SDSS J160128.54+544521.3, the presented spectra are based on $Spitzer$ IRS spectra of 5MUSES due to high redshifts instead of AKARI IRC spectra. We also include insets showing fitting results of the 3.3 $\mu m$ PAH features for individual galaxies detected with the 3.3 $\mu m$ PAH emission.

In order to detect the 3.3 $\mu m$ PAH emission feature and measure its strength, we fit the spectra around 3.3 $\micron$ using the Drude profile \citep{LD01} :

\begin{eqnarray}
\label{drudepfl}
I_{\nu}^{(r)} & = & \frac{b_{r} \gamma_{r}^{2}}{(\lambda/\lambda_{r} - \lambda_{r}/\lambda)^{2} + \gamma_{r}^{2}}, 
\end{eqnarray}
where $\lambda_{r}$ is the central wavelength of the feature, $\gamma_{r}$ is the fractional FWHM, and $b_{r}$ is the central intensity.


Fitting and subtracting continuum from spectra also can be a subject of debate. The most difficult issue is how to avoid absorption features which are abundant throughout MIR wavelength regimes. These absorption features include CO$_{2}$ ice absorption at 4.27 $\micron$, $^{13}$CO$_{2}$ ice absorption at 4.38 $\micron$, and CO ice absorption feature at 4.67 $\micron$. The vicinity of the 3.3 $\mu m$ PAH feature is not an exception with water ice absorption feature at 3.05 $\micron$ and carbonaceous dust absorption feature at 3.4 $\micron$.
Therefore, we decide to apply a spline fit for continuum to the wavelength range between 3.0 $\mu m$ and 3.6 $\mu m$ while masking out the 3.3 $\mu m$ emission feature and avoiding any obvious absorption feature by visual inspection. Generally, we used the wavelength ranges between 3.0 $\micron$ and 3.2 $\micron$, and between 3.4 $\micron$ and 3.6 $\micron$ for the continuum fit. If the 3.3 $\micron$ feature looks to be shifted upon visual inspection  (perhaps due to an unknown systematic error in wavelength calibration), we adjust the fitting ranges. However, there is no 3.3 $\micron$ feature detected with a shifted peak. Also, if there is a clear absorption feature residing within the ranges upon visual inspection, we adjust the fitting range further out from 3.3 $\micron$.

We use an IDL procedure, MPFIT, to apply the Drude fit onto the 3.3 $\mu m$ PAH feature after subtracting continuum from each individual spectrum. We do not constrain peaks and widths of Drude fit unless the IDL procedure cannot fit any obvious emission feature. We confirm the detection of the 3.3 $\micron$ PAH emission feature by visual inspection of the fitting, while considering S/N over the continuum. We measure line fluxes and EWs of the 3.3 $\mu m$ PAH emission based on the outputs of the fitting results of Drude fits and continuum fits for the sample galaxies. The flux errors are determined based on the Drude fit and its errors measured by MPFIT. These values are given in Table \ref{table2}. 

When no 3.3 $\mu m$ PAH emission is detected, we measure upper limits. In order to calculate the upper limits, first, we measure the standard deviation of data points within a wavelength range at which the 3.3 $\micron$ should be located. This range is decided by the average width obtained from the other galaxies detected with the 3.3 PAH $\micron$ emission. The average width is 17 nm. Then we assume a Drude profile which has a peak three times bigger than the standard deviation, and has a width of 17 nm which is the average width of the 3.3 $\mu m$ PAH emission detected within our sample. We take the sums under these Drude profiles as upper limits for non-detection targets.

\subsection{Stacking analysis}
\label{stack}

In order to construct representative spectra of each SED class while recovering missing 3.3 $\mu m$ features due to low S/N, we apply stacking analysis. We constructed blind-stacked spectra of the entire sample, the galaxies with the 3.3 $\mu m$ PAH feature detection, the galaxies without the 3.3 $\mu m$ PAH feature, the galaxies with AGN SEDs, and the galaxies with starburst SEDs. 

To construct the stacked spectra, we simply reproduced spectra of individual sample galaxies onto a wavelength grid set to have even step through the wavelength range between 2.5 $\mu m$ and 8 $\mu m$, then constructed the spectra for each group by simply averaging spectra of group galaxies without applying any clipping or normalization. $Spitzer$ IRS spectra provide data for any wavelength range required to construct the new spectra. We present the stacked spectra of each galaxy group in Figure \ref{fig3}.

We concern that the stacked spectra are dominated by a few galaxies with high flux. Therefore we also stacked normalized individual spectra for each group and compared the spectra to the stacked spectra produced without normalization. To do this, we normalized individual spectra by fluxes between 3.5 $\micron$ and 4.0 $\micron$, then obtained averages of them. However, overall, these stacked spectra do not show any significant departures from the ones in Figure \ref{fig3}.

\section{Results}
\label{results}

\subsection{Detection and strength of the 3.3 $\mu m$ PAH emission}
\label{line}
We detect the 3.3 $\mu m$ PAH feature from three galaxies out of 20 target galaxies. Although the 3.3 $\mu m$ PAH features are generally pretty weak across the sample, a couple galaxies of starburst SED class, namely 2MASX J16182316+5527217 and 2MASX J16205879+5425127 show very clear and strong 3.3 $\mu m$ PAH features. While these two galaxies have starburst SEDs, the remaining one galaxy, 2MASX J10542172+5823445 has an AGN SED. Therefore, the detection rate for the starburst SED class sample is 29\% (2/7), while the detection rate for the AGN SED class sample is 8\% (1/12). However, since the desired S/N is not achieved for a large portion of the sample, these detection rates should be considered as lower limits. 

Upper limits varies significantly. However, generally it is smaller for the starburst SED galaxies than the AGN SED galaxies. While the five SB galaxies without the 3.3 $\micron$ PAH emission detection have their average upper limits of 5.97$\times 10^{41}$ erg s$^{-1}$, their AGN counterparts have detection limits of  1.35$\times 10^{43}$ erg s$^{-1}$.

The 3.3 $\micron$ PAH luminosities range from 7.11 $\times$10$^{42}$ erg s$^{-1}$ of an AGN at z = 0.205 (2MASX J10542172+5823445) to 7.14 $\times 10^{41}$ erg s$^{-1}$ of a starburst galaxy at z =0.063 (2MASX J16205879+5425127). The two starburst galaxies with 3.3 $\micron$ PAH detection are at z $\sim$ 0.1 and their average luminosity is 9.46 $\times 10^{41}$ erg s$^{-1}$.

\subsection{Stacked spectra}
\label{stackresult}

The stacked spectrum of the entire sample shows a distinctive emission feature of the 3.3 $\mu m$ PAH emission. The EW of the stacked spectrum is 11 nm, while the spectrum stacked with the galaxies with starburst SEDs has the EW of 32 nm. On the other hand, the spectrum stacked with the non-detection galaxies does not reveal the 3.3 $\mu m$ PAH emission feature. The 3.3 $\mu m$ PAH emission does not show up for the stacked AGN spectrum, either. It is quite obvious that AGN host galaxies do not have strong PAH emission overall.

\section{Discussion}
\label{discussion}

\subsection{The Correlation between $L_{\mathrm{PAH3.3}}$ and $L_{\mathrm{IR}}$}
\label{trend}

We present a plot comparing luminosity of the 3.3 $\mu m$ PAH emission ($L_{\mathrm{PAH3.3}}$) with infrared luminosity, $L_{IR}$ (Figure \ref{fig4}). Due to lack of the detected sources within the AMUSES sample, it is not meaningful to derive the correlation between $L_{\mathrm{PAH3.3}}$ and $L_{\mathrm{IR}}$ only for the AMUSES sample. Therefore we look for the correlation for the combined sample of AMUSES and literature samples. Big filled circles represent samples of the AMUSES objects, while small symbols represent samples from literatures \citep{RAV03, Im08, Sj09, Im10, LJC}. Note that $L_{\mathrm{IR}}$ come from \citet{5muses}. In order to estimate $L_{\mathrm{IR}}$ of the sample of 5MUSES, \citet{5muses} utilize mid-IR spectra from IRS along with the IRAC and MIPS photometry. First, they construct an IR template library based on these data. This library covers a wide range of galaxies from normal star-forming galaxies to ULIRGs to quasars. Then using synthetic IRAC photometry drawn from the IRS spectra and MIPS photometry, they compare them with the corresponding synthetic photometry from the SED templates and estimates $L_{\mathrm{IR}}$. 
On the other hand, the other samples utilize IRAS photometry to calculate $L_{\mathrm{IR}}$. More specifically, they used the formula given in Table 1 of \citet{SM96}.


The figure shows a broad correlation between $L_{\mathrm{PAH3.3}}$ and $L_{\mathrm{IR}}$ for the combined sample of our data points and the data from the literature, though the scatter in the relation is large. In order to obtain a linear correlation between $L_{\mathrm{PAH3.3}}$ and $L_{\mathrm{IR}}$, we carry out $\chi^{2}$ fitting to the combined sample while switching abscissa and ordinate between $L_{\mathrm{PAH3.3}}$ and $L_{\mathrm{IR}}$, since the slope of the correlation changes significantly (by 15 \%) depending on which quantity is chose as a dependent. Then we take the average fit as our fitting result. The linear fit to the combined sample is as in Eq. (\ref{LIRL33});
\begin{eqnarray}
\label{LIRL33}
\mathrm{log} (L_{\mathrm{IR}})& = & (1.16 \pm 0.30) \times \mathrm{log} (L_{\mathrm{PAH3.3}}) \nonumber \\
& & -(3.11 \pm 0.34), 
\end{eqnarray}
where $L_{\mathrm{IR}}$ and $L_{\mathrm{PAH3.3}}$ are in the unit of erg sec$^{-1}$ with the correlation coefficient of 0.70. This fit is shown in Figure \ref{fig4} with a solid line. The undetected sources are not included for this fitting. In general, the data points of AMUSES sample show higher $L_{\mathrm{PAH3.3}}$ than the literature sample. Considering the fact that we may have a rather shallow detection limit and an unbiased distribution for AMUSES sample, the detected sources of AMUSES sample are more luminous objects in the 3.3. $\micron$ PAH emission feature and occupy the upper envelope in terms of $L_{\mathrm{PAH3.3}}$ for the correlation between $L_{\mathrm{PAH3.3}}$ and $L_{\mathrm{IR}}$ of the literature samples. 

On the other hand, $L_{\mathrm{IR}}$ can contain non-star-forming contribution. Therefore, we consider the correlation between $L_{\mathrm{PAH3.3}}$ and $L_{\mathrm{IR}}$ only for sources with starburst SEDs, or HII-like SEDs. The fit is given as in Eq. (\ref{LIRL33sb});
\begin{eqnarray}
\label{LIRL33sb}
\mathrm{log} (L_{\mathrm{IR}})& = & (1.37 \pm 0.17) \times \mathrm{log} (L_{\mathrm{PAH3.3}}) \nonumber \\ &&-(12.18 \pm 0.75), 
\end{eqnarray}
where the units for $L_{\mathrm{PAH3.3}}$ and $L_{\mathrm{IR}}$ are same with Eq. (\ref{LIRL33}). The correlation coefficient for the fit is 0.69. We use the same method which we use for Eq. (\ref{LIRL33}) for this fit. This fit is shown within Figure \ref{fig4} by a dotted line.

\subsection{The PAH 3.3 $\micron$ emission as a star formation indicator}
\label{sf}

In order to to be an effective star formation indicator, $L_{\mathrm{PAH3.3}}$ must have a good correlation with $L_{\mathrm{IR}}$. Within the detected sources from AMUSES sample and the literature samples, especially with the sources with SB SEDs, $L_{\mathrm{PAH3.3}}$ and $L_{\mathrm{IR}}$ have a correlation between them, but with a large scatter (Figure \ref{fig4}).

This can be also confirmed by previous studies on other PAH emission features. Although systematic calibration of any particular PAH emission feature to SFR has been carried out less extensively than other SFR indicators, there have been a few studies to show strong correlations between the strengths of PAH emission features and IR luminosity. \citet{Gz98} utilizes the the strength of the 7.7 $\micron$ PAH emission for diagnosing energy sources for ULIRGs based on 15 spectra from ISO. Also using ISO spectra of galactic and extragalactic star sources, \citet{Pt04} shows that the luminosity of the 6.2 $\micron$ PAH emission ($L_{\mathrm{PAH6.2}}$) correlates very well with L$_{FIR}$. \citet{Br06} calibrate the 6.2 $\micron$ PAH emission to $L_{\mathrm{IR}}$ using IRS spectra of 22 starburst galaxies.
Regarding more of higher redshift objects, \citet{Pp08} found that all the PAH bands which they detected from 13 sub-mm galaxies (SMGs) at various redshifts between 1.0 and 2.5, namely the 6.2, 7.7, and 11.3 $\micron$ PAH emission features, correlate well with $L_{\mathrm{IR}}$. 
\citet{Shi09} also found that the 6.2 $\micron$ and 7.7 $\micron$ PAH emission features correlate well with SFR based on $L_{\mathrm{IR}}$ and use them to measure the cosmic star formation history of QSOs. \citet{Ft10} shows that the 6.2 $\micron$ and 7.7 $\micron$ PAH luminosities of their 16 ULIRGs at z $\sim$ 2 follow the correlation of the samples of \citet{Pp08}, \citet{Md09}, and \citet{Shi09}. On top of these studies, \citet{5muses} found strong correlations between $L_{\mathrm{PAH6.2}}$ and $L_{\mathrm{IR}}$ for SB, composite, and AGN sources based on 5MUSES, the source of the parent sample for this study (A1, A2, and A3 of \citet{5muses}).

Whether our $L_{\mathrm{PAH3.3}}$-$L_{\mathrm{IR}}$ correlation persists, especially at high $L_{\mathrm{IR}}$ remains as a subject of further investigation.
In Figure \ref{fig6}, we plot $L_{\mathrm{IR}}$ against $L_{\mathrm{PAH6.2}}$. We adopt $L_{\mathrm{PAH6.2}}$ from \citet{5muses}. 
\citet{5muses} utilize the PAHFIT software \citep{Sm07} and measured $L_{\mathrm{PAH6.2}}$ from the fitting result based on it.
A linear fit to the detections of AMUSES sample  gives the correlation between $L_{\mathrm{PAH6.2}}$ and $L_{\mathrm{IR}}$ as in Eq. (\ref{LIRL62});
\begin{eqnarray}
\label{LIRL62}
\mathrm{log} (L_{\mathrm{IR}})& = & (0.95 \pm 0.03) \times \mathrm{log} (L_{\mathrm{PAH6.2}}) \nonumber \\ && +(4.21 \pm 0.19), 
\end{eqnarray}
where the units for $L_{\mathrm{PAH6.2}}$ and $L_{\mathrm{IR}}$ are same with Eq. (\ref{LIRL33}) with the correlation coefficient of 0.75.
However, it is noticeable that the ULIRGs deviate from the correlation. They have either higher $L_{\mathrm{IR}}$, or lower $L_{\mathrm{PAH6.2}}$ than lower $L_{\mathrm{IR}}$ objects suggest. It looks plausible that ULIRGs at $L_{IR} > 10^{12} M_{\odot}$ may deviate from correlations between PAH luminosity and $L_{\mathrm{IR}}$, since these ULIRGs also may deviate from the correlation between $L_{\mathrm{PAH3.3}}$ and $L_{\mathrm{IR}}$ (Figure \ref{fig4}). Similar trends are reported by \citet{Pt04} and \citet{Im07}. In order to check if this is true, we obtain a linear fit to the detections with SB SEDs which are not ULIRGs. This fit gives the correlation between $L_{\mathrm{PAH3.3}}$ and $L_{\mathrm{IR}}$ as in Eq. (\ref{LIRL33noul});
\begin{eqnarray}
\label{LIRL33noul}
\mathrm{log} (L_{\mathrm{IR}})& = & (0.82 \pm 0.13) \times \mathrm{log} (L_{\mathrm{PAH3.3}}) \nonumber \\ &&+(10.58 \pm 0.86), 
\end{eqnarray}
where the units for $L_{\mathrm{PAH3.3}}$ and $L_{\mathrm{IR}}$ are same with Eq. (\ref{LIRL33}). The correlation coefficient for the fit is 0.71 and the fit is shown in Figure \ref{fig4} as a dashed line.

It is noticeable that ULIRGs generally have smaller $L_{\mathrm{PAH3.3}}$ for a given $L_{\mathrm{IR}}$ regardless of SED classes based on this fit. Similar trends are found by \citet{Cl00} for the 7.7 $\micron$ PAH emission feature and \citet{Pt04} for the 6.2 $\micron$ PAH emission feature. It is not plausible that the offset for ULIRGs are due to any systematic difference in measurement methods for $L_{\mathrm{PAH3.3}}$, or $L_{\mathrm{IR}}$. For example, it is unlikely if $L_{\mathrm{IR}}$ based on IRAS four band photometry overestimates true $L_{\mathrm{IR}}$, or $L_{\mathrm{IR}}$ which we adopted from \citet{5muses} underestimates. 

There can be two possible explanations why ULIRGs show the trend. First, ULIRGs may have larger non-star-forming contribution to their $L_{\mathrm{IR}}$ regardless of their SED classes based on their optical spectra. There have been studies claiming that AGN contribution to $L_{\mathrm{IR}}$ is not negligible for ULIRGs (\citealt{ntz09, So10}, and references therein). For instance, \citet{lz98} show that, while only 15\% of ULIRGs at luminosities below
2 $\times$ 10$^{12}$ L$_{\odot}$ are attributed to AGN, this fraction increases to 50\% at higher luminosity. Figure 5 of \citet{ntz07} show the correlations between nuclear activity and two IR luminosities; FIR(60$\micron$) luminosity (L$_{60}$) and the 7.7 $\micron$ PAH luminosity. Interestingly, the correlation between the nuclear activity probed by the $\lambda$5100$\AA$ monochromatic luminosity and the 7.7 $\micron$ PAH luminosity breaks at about L$_{60}$ $\sim$ 10$^{12}$ L$_{\odot}$, while L$_{60}$ correlates well with the entire range of $\lambda$5100$\AA$ monochromatic luminosity. The difference between these two correlations support that AGN activity may contribute to $L_{\mathrm{IR}}$ more than it has been considered to be. Also, considering the fact that there is no sign of departure for ULIRG sample of \citet{Im08} from the correlation between $L_{\mathrm{PAH3.3}}$ and $L_{\mathrm{PAH6.2}}$ (Figure \ref{fig5}), it is more likely to be $L_{\mathrm{IR}}$ which causes the break. Recently, \citet{YR} found that the correlation between $L_{\mathrm{PAH3.3}}$/$L_{\mathrm{IR}}$ and $L_{\mathrm{IR}}$ has a break around $L_{\mathrm{IR}}$ $\sim$ 10$^{12}$ L$_{\odot}$ from another MP of AKARI, MSAGN. There is another candidate to contribute to $L_{\mathrm{IR}}$ for ULIRGs other than AGN activity: embedded young stellar objects (YSOs) \citep{Pt04, Sp02, Sp07}. Although embedded YSOs are from star-forming regions, they should be considered as a non-typical star-forming contribution to $L_{\mathrm{IR}}$. Based on ISO observation, \citet{Sp02} found that most ULIRGs have the 6 $\sim$ 8 $\micron$ ice absorption feature and linked it to strongly obscured beginning of star formation. 

Second, ULIRGs suppress PAH emissions, not only the 3.3 $\micron$ PAH emission, but the PAH emission features in general. As shown in Figure \ref{fig4} and Figure \ref{fig6}, both $L_{\mathrm{PAH3.3}}$ and $L_{\mathrm{PAH6.2}}$ are lower for ULIRGs for a given $L_{\mathrm{IR}}$. It is believed that strong UV radiation, such as one from AGN, can destruct PAH molecules and reduce the population of PAHs. In fact, \citet{Im08} claimed that the ratio between the 3.3 $\micron$ PAH luminosity and infrared luminosity for ULIRGs are less than what is expected from obscured starbursts and are attributed to the deficiency of PAH within ULIRGs. However, it is unclear why AGN at L$_{IR} < 10^{12}$ L$_{\odot}$ does not reduce the PAH luminosity as much. Therefore, it is plausible that most ULIRGs have a large contribution from AGN to their $L_{\mathrm{IR}}$. A multi-wavelength analysis of the ULIRGs, including the mid-IR component from hot dust, should settle this issue \citep{Nd09, Nd10, Sj12}.

However, it is still possible that there is no such departure from the correlations for ULIRGs. Since there are few sources detected with the PAH emission features at the lower end, the correlations between the PAH luminosities and $L_{\mathrm{IR}}$ can be skewed by the few objects with large non-star-forming contribution to their $L_{\mathrm{IR}}$. In order to see if it is the case, it is required to detect more lower luminous objects in terms of PAH luminosity.

On the other hand, as shown in Figure \ref{fig4} and Figure \ref{fig6}, the 3.3 $\micron$ and 6.2 $\micron$ emission features behave similarly against $L_{\mathrm{IR}}$. In order to see if there is any difference between them, we plot $L_{\mathrm{PAH3.3}}$ against $L_{\mathrm{PAH6.2}}$ and present it in Figure \ref{fig5}. Although it looks to have large scatters, the 3.3 $\micron$ PAH emission has a strong correlation with the 6.2 $\micron$ PAH emission. A linear fit to the data points of all samples without the upper limits gives the correlation between $L_{\mathrm{PAH3.3}}$ and $L_{\mathrm{PAH6.2}}$ as in Eq. (\ref{L33L62});
\begin{eqnarray}
\label{L33L62}
\mathrm{log} (L_{\mathrm{PAH6.2}})& = & (0.83 \pm 0.06) \times \mathrm{log} (L_{\mathrm{PAH3.3}}) \nonumber \\ &&+ (7.88 \pm 0.41), 
\end{eqnarray}
where the units for $L_{\mathrm{PAH3.3}}$ and $L_{\mathrm{PAH6.2}}$ are erg sec$^{-1}$ with the correlation coefficient of 0.83. This fit is shown within Figure \ref{fig5} by a solid line.
Interestingly, ULIRGs do not deviate from the overall trend. These ULIRGs deviate rather significantly from the correlation between $L_{\mathrm{IR}}$ and PAH luminosities (Figure \ref{fig4} and Figure \ref{fig6}). Therefore, it assures that the 3.3 $\micron$ PAH can represent what the 6.2 $\micron$ represents, if needed. However, it is noticeable that the correlation between $L_{\mathrm{PAH3.3}}$ and $L_{\mathrm{PAH6.2}}$ is far from unity. This may be attributed to the origin of PAH emission features. As mentioned before, \citet{vD04} compare variations of the profiles of various PAH emission features and show that while the 6.2 and 7.7 $\micron$ emission originate from PAH cations, the 3.3 $\micron$ emission feature is attributed to neutral and/or negatively charged PAHs, which is suggested by several laboratory studies \citep{SV93, Lf96,HA99}.

Therefore, although the 3.3 $\micron$ PAH emission feature shows that it has the potential to be a SFR proxy like other PAH emission features, it heavily relies on understanding the physical conditions of PAH molecules and their emission features. It also requires to understand the deviation of ULIRGs and its origin lest the power of $L_{\mathrm{PAH3.3}}$ as a proxy is hampered by this deviation.

\section{Summary}
\label{sum}

We describe the details of a mission project of AKARI, AMUSES and its results. With AMUSES, we have investigated the 3.3 $\mu m$ PAH feature and its correlation with other PAH features as well as other properties of the sample galaxies. The summary of our results are:

\begin{enumerate}

\item{We detected the 3.3 $\mu m$ PAH emission from three out of 20 flux limited sample galaxies and the detection rate for our sample is 15\%.}

\item{For the combined sample of AMUSES and the literature samples, $L_{\mathrm{PAH3.3}}$ correlates with $L_{\mathrm{IR}}$, while $L_{\mathrm{PAH6.2}}$ also has a similar correlation with $L_{\mathrm{IR}}$.}

\item{ULIRGs at $L_{IR} > 10^{12} L_{\odot}$ may deviate from the correlation between $L_{\mathrm{PAH3.3}}$ and $L_{\mathrm{IR}}$ of the lower luminous objects due to non-star-forming contribution on $L_{\mathrm{IR}}$, such as AGN activity and heavily obscured YSOs, or destruction of PAH molecules by AGN activity.}

\end{enumerate}

{\bf Acknowledgement}
This work was supported by Korean Research Foundation (KRF) grant funded by the Korean government (MEST), No. 2010--0000712.
This research is based on observations with AKARI, a JAXA project with the participation of ESA.


\begin{deluxetable}{cccccccc}
\tabletypesize{\scriptsize}
\tablecaption{Basic Properties of Sample Galaxies
\label{table1}
}
\tablewidth{0pt}
\tablehead{
\colhead{Galaxy} 
&\colhead{RA} &\colhead{Dec}
&\colhead{redshift}
&\colhead{$f_{3.6}^{\dagger}$}
&\colhead{SED class}
&\multicolumn{2}{c}{number of pointings}\\
\colhead{}
&\colhead{}
&\colhead{}
&\colhead{}
&\colhead{(mJy)}
&\colhead{}
&\colhead{NG}
&\colhead{NP}
}

\startdata
FBQS J0216-0444  & 2:16:40.72 & -4:44:05.1 & 0.87 & 2.11& AGN & 3 & 1\\
2MASX J02165778-0324592  & 2:16:57.77 & -3:24:59.7 & 0.137 & 2.53& AGN &1 & 0\\
2MASX J02191605-0557269   & 2:19:16.05 & -5:57:26.9 & 0.103 & 1.04& AGN & 3 &1\\
2MASX J02193906-0511336   & 2:19:30.08 & -5:11:33.8 & 0.151 &1.93 & AGN &1 &0\\
2MASX J02195305-0518236   & 2:19:53.04 & -5:18:24.1 & 0.072 & 1.40& SB & 1 &0\\
2MASX J10520659+5809476   & 10:52:06.56 & +58:09:47.1 & 0.1172 &1.34 & SB & 1 &0\\
2MASX J10542172+5823445   & 10:54:21.65 & +58:23:44.6 & 0.2045 & 2.58& AGN &2 &0 \\
VII ZW353   & 10:57:05.43 & +58:04:37.4 & 0.1404 & 2.35& AGN &3 &1\\
SDSS J105959.93+574848.1   & 10:59:59.95 & +57:48:48.1 & 0.4530 &1.20 & AGN & 3 &0\\
SDSS J155936.13+544203.8    & 15:59:36.12 & +54:42:03.7 & 0.3077 & 1.09& AGN & 3 &1\\
SDSS J160128.54+544521.3    & 16:01:28.52 & +54:45:21.3 & 0.7278 & 1.13 & AGN & 3 &1\\
2MASX J16044063+5534089  & 16:04:40.64 & +55:34:09.2 & 0.078 & 1.23& SB &2 & 0\\
2MASX J16130186+5521231   & 16:13:01.82 & +55:21:23.0 & 0.012 & 2.07& SB & 1 &0\\
2MASX J16144902+5545120   & 16:14:45.92 & +55:45:12.9 & 0.064 & 1.55& AGN & 2 &0\\
SBS1614+546   & 16:15:21.78 & +54:31:48.3 & 0.474 & 1.18& AGN &1 & 0\\
SDSS J161445.94+542554.4    & 16:16:45.92 & +54:25:54.4 & 0.223 & 1.06& AGN &1 &0\\
2MASX J16165997+5600276   & 16:16:59.95 & +56:00:27.2 & 0.063 & 2.35& SB & 3 &1\\
2MASX J16181934+5418587  & 16:18:19.31 & +54:18:59.0 & 0.083 & 1.50& Composite & 2 &0\\
2MASX J16182316+5527217   & 16:18:23.11 & +55:27:21.4 & 0.084 & 1.71& SB &3 & 1\\
2MASX J16205879+5425127   & 16:20:58.82 & +54:25:13.1 & 0.082 & 1.71& SB & 3 &1

\enddata

\tablecomments{$\dagger$ $Spizter$ IRAC 1 band fluxes from 5MUSES \citep{5muses}.}

\end{deluxetable}

\begin{deluxetable}{ccccccc}
\tabletypesize{\scriptsize}
\tablecaption{PAH measurement of the sample galaxies
\label{table2}
}
\tablewidth{0pt}
\tablehead{
\colhead{Galaxy} 
&\colhead{$f_{3.3}^{\dagger}$} 
&\colhead{EW$_{3.3}(nm$)}
&\colhead{$f_{6.2}^{\ddagger}$}
&\colhead{$f_{7.7}^{\ddagger}$}
&\colhead{$f_{11.3}^{\ddagger}$}
&\colhead{$log$ LIR(M/M$_{\odot}$)}
}

\startdata
FBQS J0216-0444  & $<$589.9 & $<$10.4 & 2.407  & .... & 15.23  & 12.70 $\pm$ 0.01 \\
2MASX J02165778-0324592  & $<$4320.3& $<$38.2 & 1.108  & 6.167 & 10.04  & 10.90 $\pm$0.03\\
2MASX J02191605-0557269   & $<$1853.6 & $<$9.3 & 14.57  & 59.89 & 33.77  & 10.71 $\pm$ 0.05 \\
2MASX J02193906-0511336   & $<$3268.3 & $<$49.8 &  44.66 & 230.2 &  72.83 & 11.38 $\pm$0.06 \\
2MASX J02195305-0518236   & $<$3504.7 & $<$17.1 & 292.7  & 931.5 & 253.5  & 10.93 $\pm$0.03 \\
2MASX J10520659+5809476   & $<$5290.1 &$<$3.2 &  231.3 & 821.1 & 228.6  & 11.34 $\pm$ 0.03 \\
2MASX J10542172+5823445   & 5882.3$\pm$2083.9 & 115.2$\pm$40.8 &  50.21 & 222.8 & 65.59  & 11.43 $\pm$ 0.03 \\
VII ZW353   & $<$1703.0 & $<$4.9 &  40.10 & 220.6 & 92.47 & 11.18 $\pm$ 0.03 \\
SDSS J105959.93+574848.1   & $<$2873.6 &$<$32.3 & 5.993  & 20.28 & 11.79  & 11.83 $\pm$ 0.02 \\
SDSS J155936.13+544203.8    & $<$2795.4& $<$56.7 &  6.254 & 13.40 & 22.42 & 11.32 $\pm$ 0.06 \\
SDSS J160128.54+544521.3    & $<$794.4 & $<$18.5 &  7.008 & 17.24 & 41.84  & 12.47 $\pm$ 0.01 \\
2MASX J16044063+5534089  & $<$ 2608.4  & $<$ 4.5  & 152.5  & 559.9 & 173.7  & 11.10 $\pm$ 0.04 \\
2MASX J16130186+5521231   & $<$6279.6 & $<$10.5 & 291.1   & 1045 & 338.4  & 9.47 $\pm$ 0.05 \\
2MASX J16144902+5545120   & $<$3712.7 &$<$7.2 &  46.75 & 103.7 & 35.10  & 10.26 $\pm$ 0.03  \\
SBS1614+546   & $<$7860.7 & $<$125.3 & 5.825  & 16.20 & 12.97  & 11.47 $\pm$ 0.08 \\
SDSS J161445.94+542554.4    & $<$3484.4 & $<$43.9 &  33.79 & 88.38 & 45.77  & 11.26 $\pm$ 0.02 \\
2MASX J16165997+5600276   & $<$ 2574.6 & $<$ 2.2 &  165.0 & 597.8 & 185.5  & 10.66 $\pm$ 0.02 \\
2MASX J16181934+5418587  & $<$3975.1 & $<$4.8 &  173.4 & 685.4 & 193.6  & 11.14 $\pm$ 0.04 \\
2MASX J16182316+5527217   & 6734.2$\pm$2053.4 & 170.1$\pm$51.9 &   435.4 & 1456 & 398.5  & 11.13 $\pm$ 0.03 \\
2MASX J16205879+5425127   & 4290.7$\pm$1807.3 & 67.3$\pm$28.3 &  297.6  & 954.9 & 259.1  & 11.11 $\pm$ 0.03
\enddata

\tablecomments{ ${\dagger}$ Line fluxes of the 3.3 $\micron$ PAH feature with units of 10$^{-17}$erg s$^{-1}$cm$^{-2}$. $\ddagger$ Line fluxes of the other PAH feature from \citet{5muses} with units of 10$^{-15}$ erg s$^{-1}$cm$^{-2}$.
}
\end{deluxetable}

\begin{figure}
\begin{center}
\includegraphics[angle=90, scale=0.7]{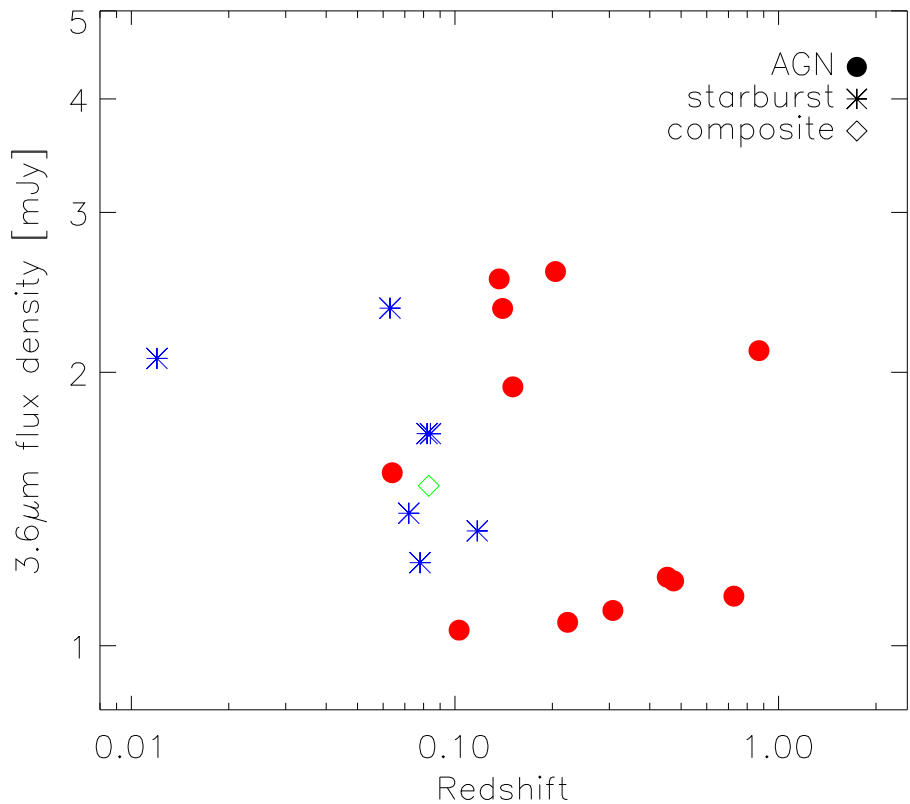}
\caption{Redshift distribution of the sample of galaxies of AMUSES plotted against 3.6 $\mu m$ flux. Symbols represent the SED classification of target galaxies based on their MIR colors by 5MUSES collaboration. Filled circles represent the AGN-type SED, while asterisks represent the starburst-type SED. An open diamond represents the composite SED target. AGN-type targets are more evenly distributed across the redshift range, while starburst-type targets are more clustered at lower redshifts.
\label{fig1}}
\end{center}
\end{figure}

\begin{figure}
\begin{center}
\includegraphics[angle=90, scale=0.35]{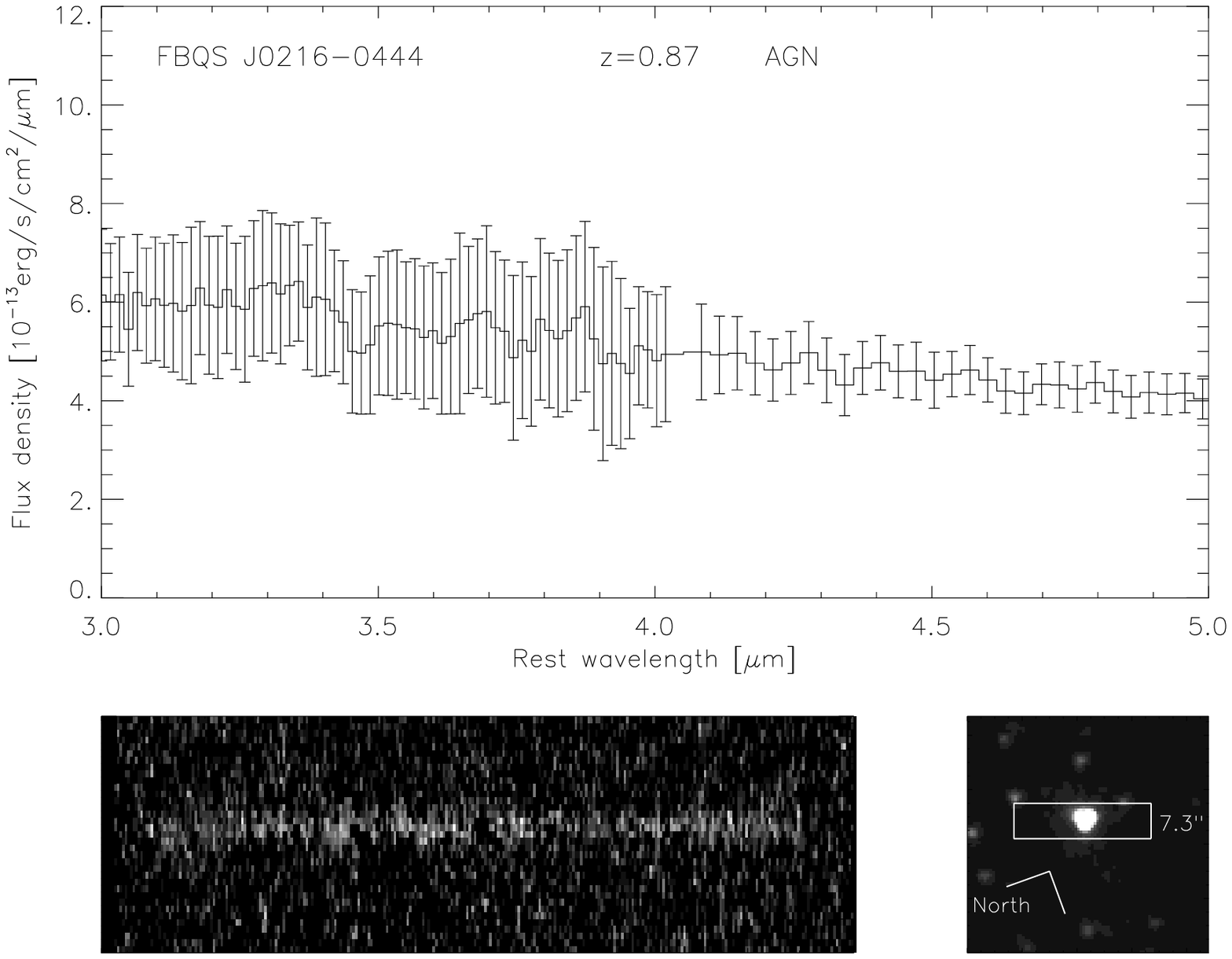}
\includegraphics[angle=90, scale=0.35]{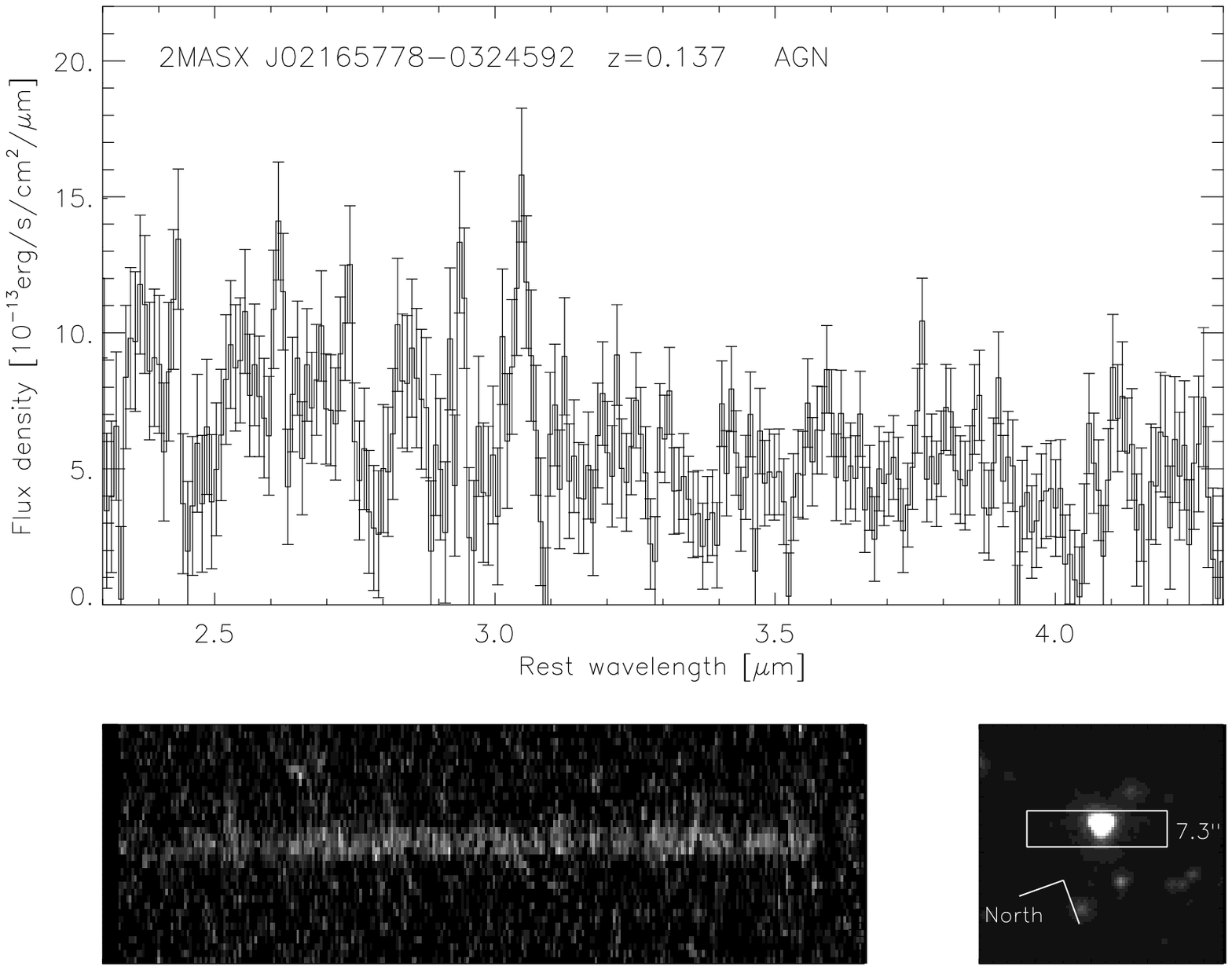}
\includegraphics[angle=90, scale=0.35]{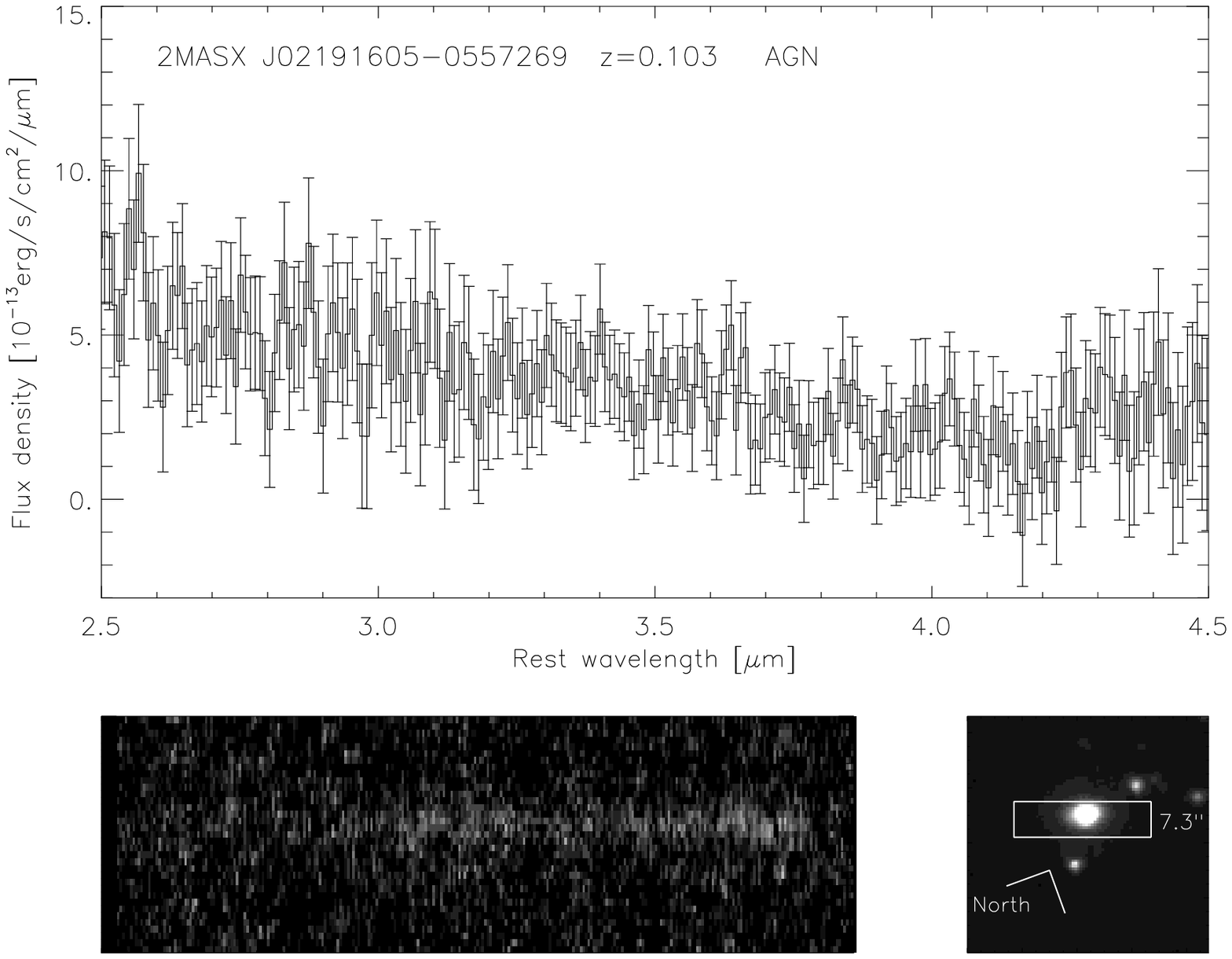}
\includegraphics[angle=90, scale=0.35]{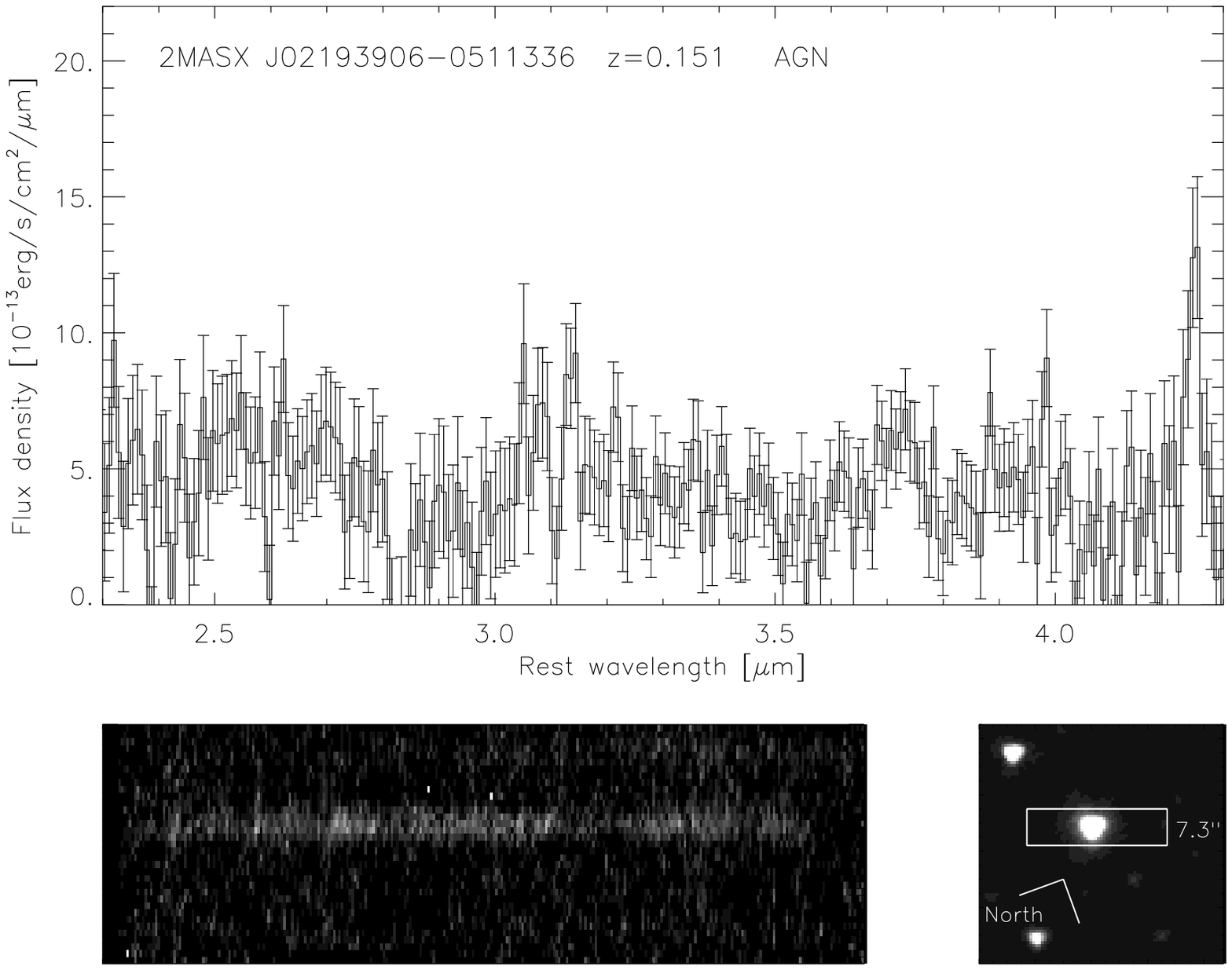}
\includegraphics[angle=90, scale=0.35]{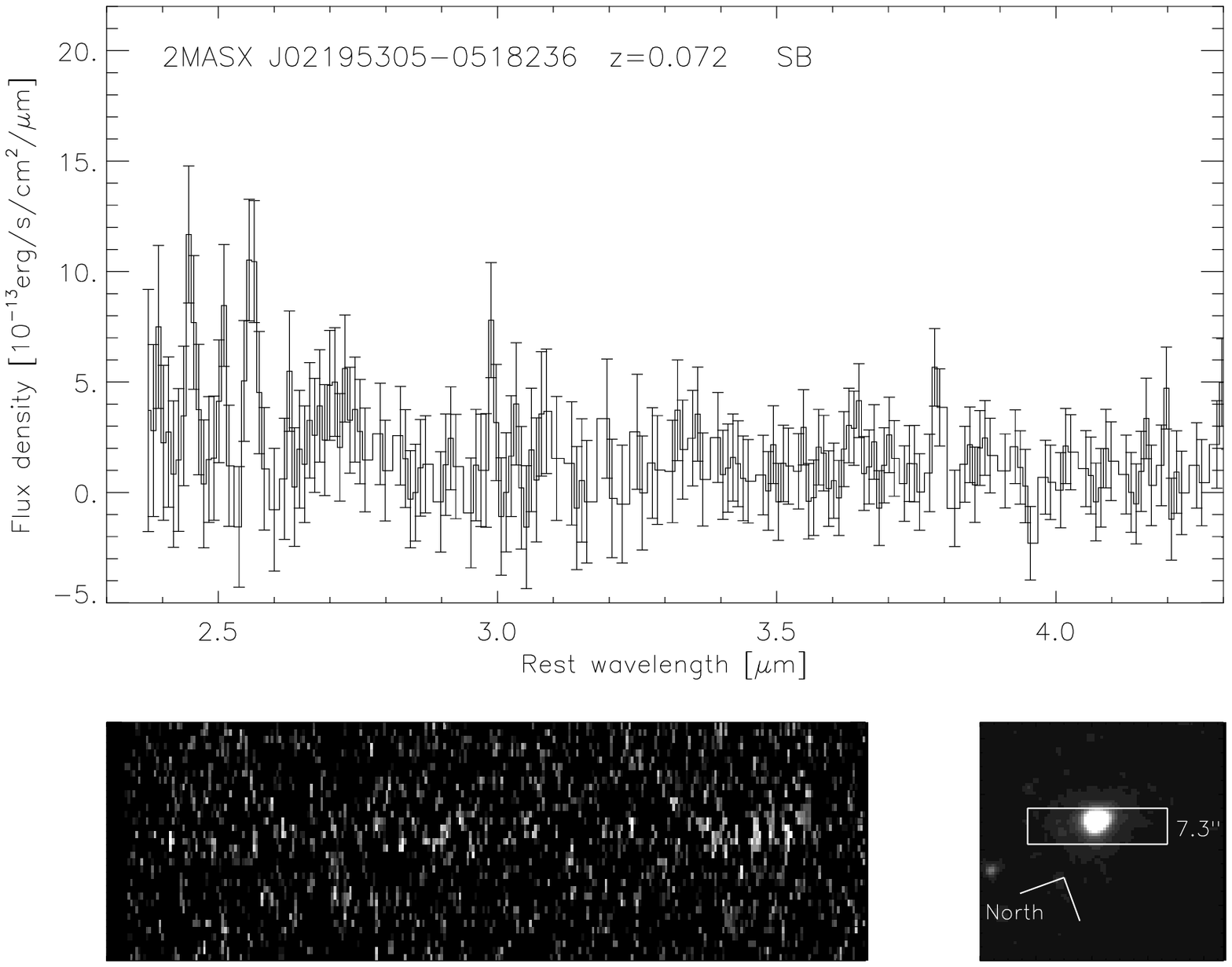}
\includegraphics[angle=90, scale=0.35]{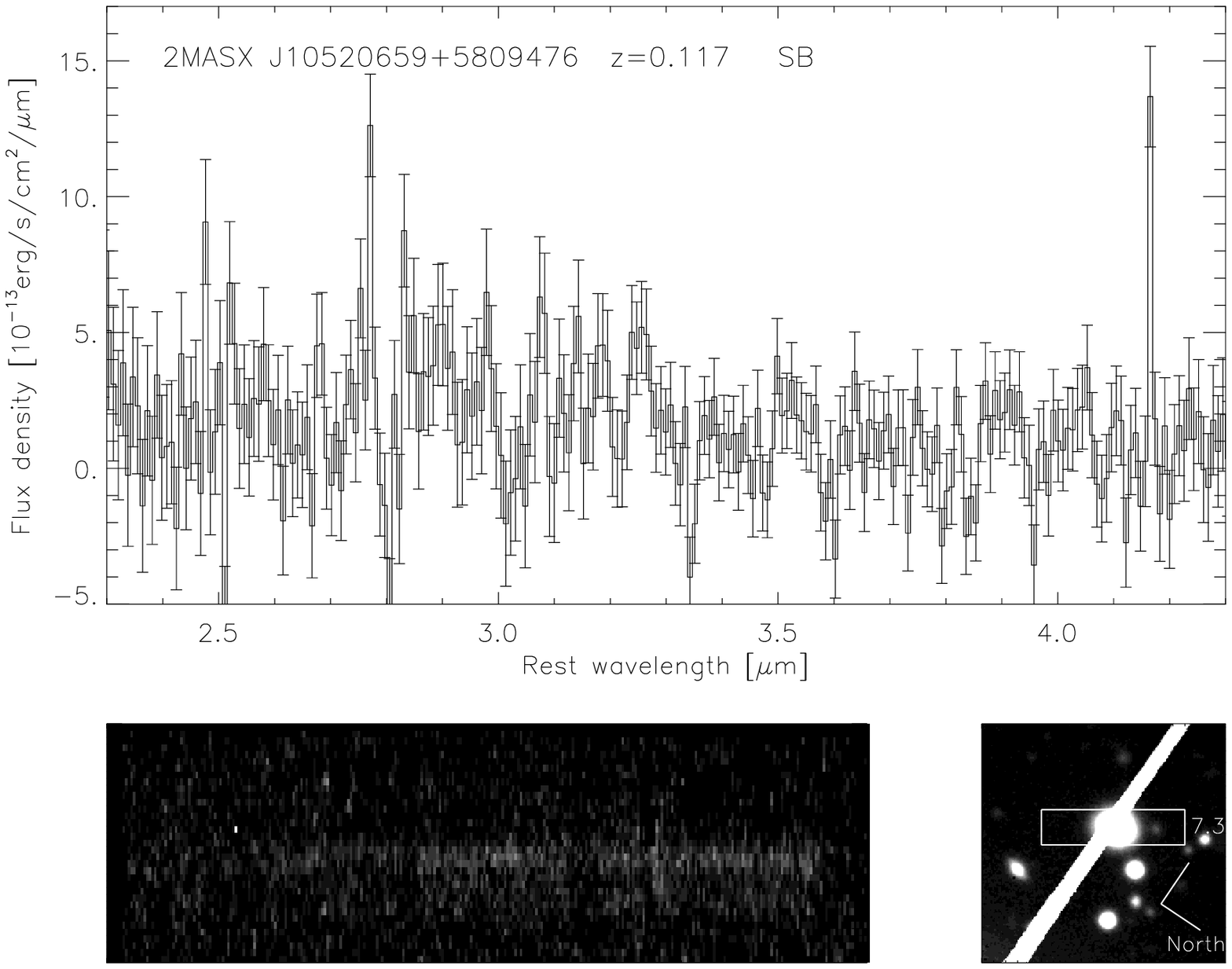}
\includegraphics[angle=90, scale=0.35]{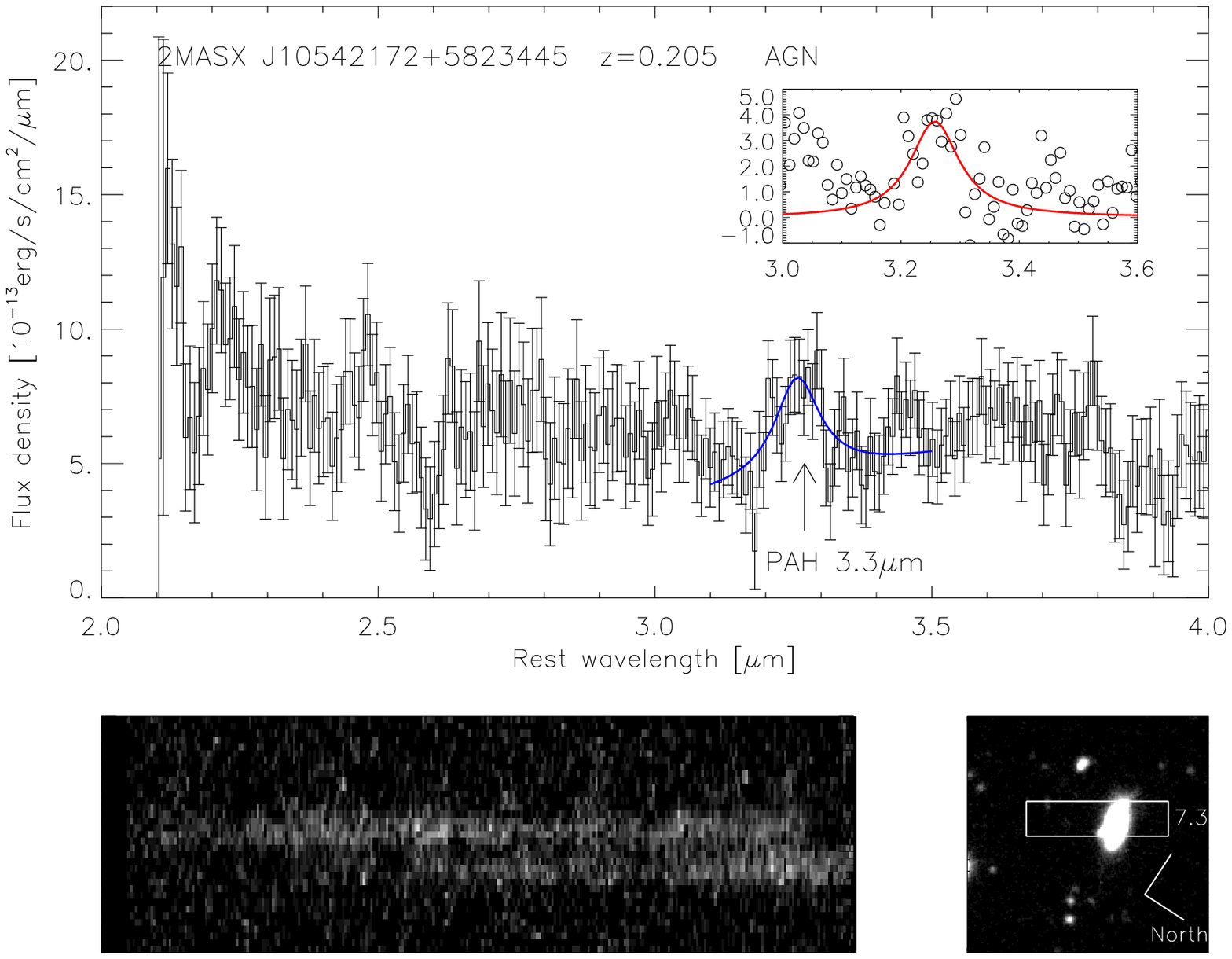}
\includegraphics[angle=90, scale=0.35]{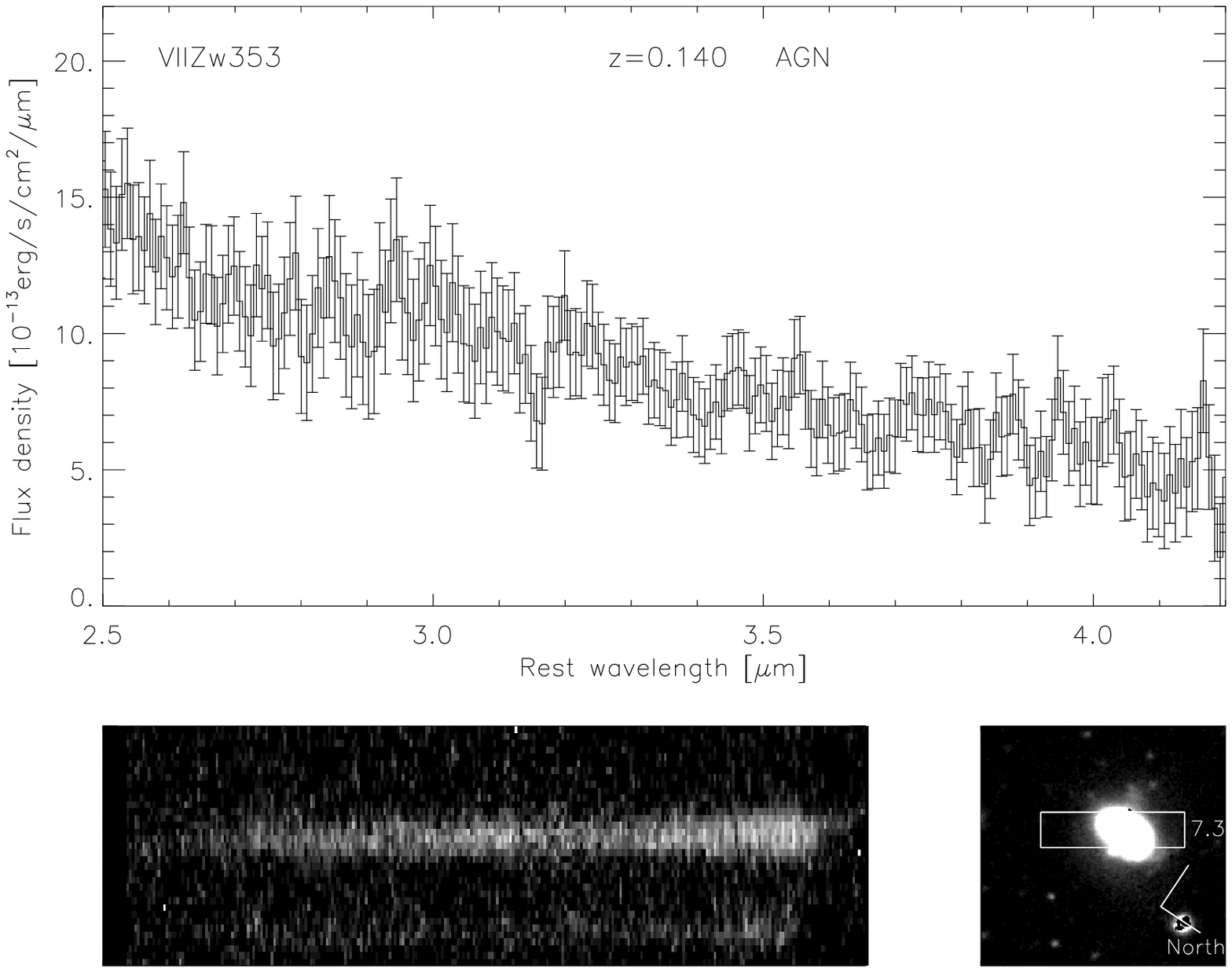}
\caption{Continued.
\label{fig2}}
\end{center}
\end{figure}

\begin{figure}
\figurenum{2}
\begin{center}
\includegraphics[angle=90, scale=0.35]{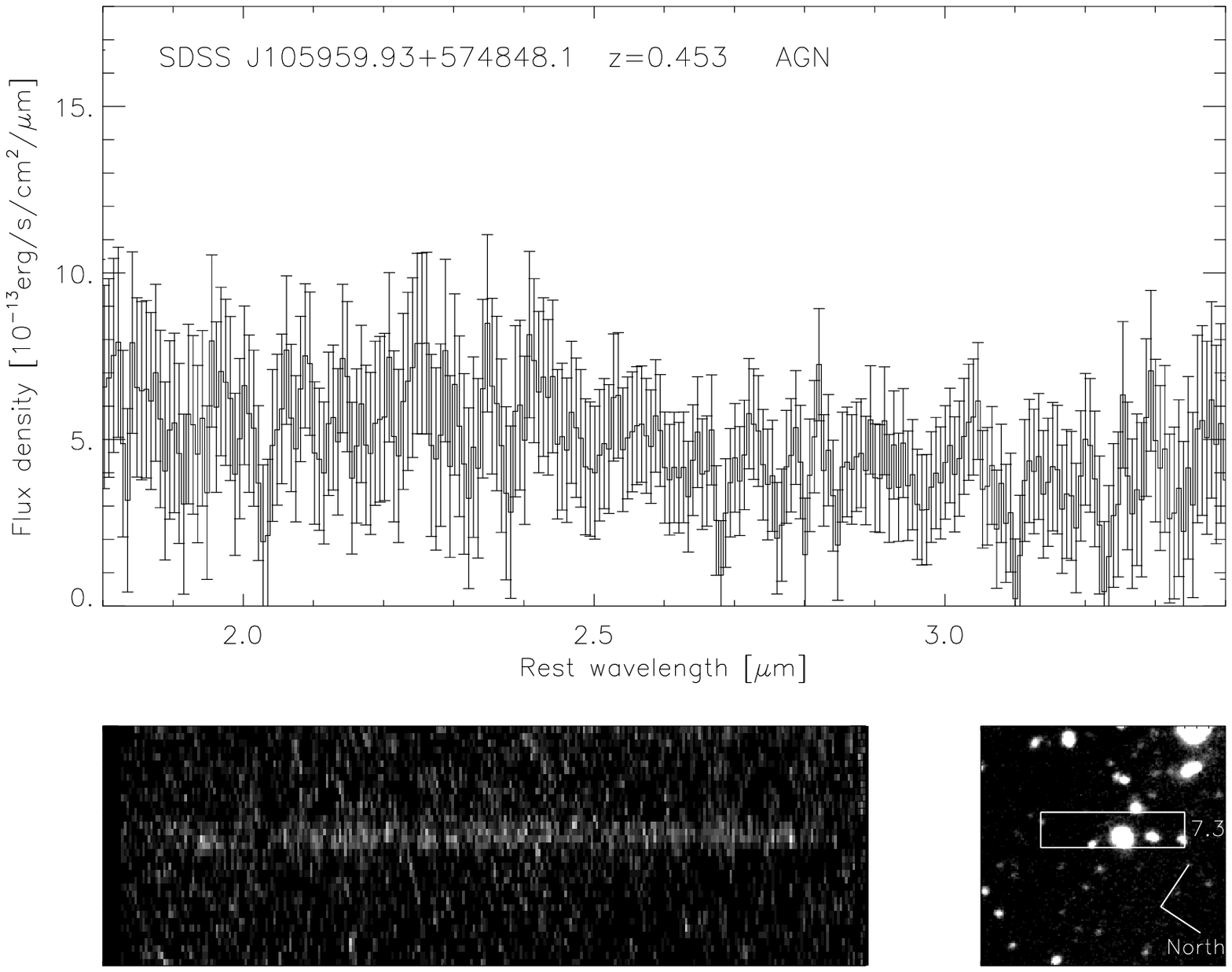}
\includegraphics[angle=90, scale=0.35]{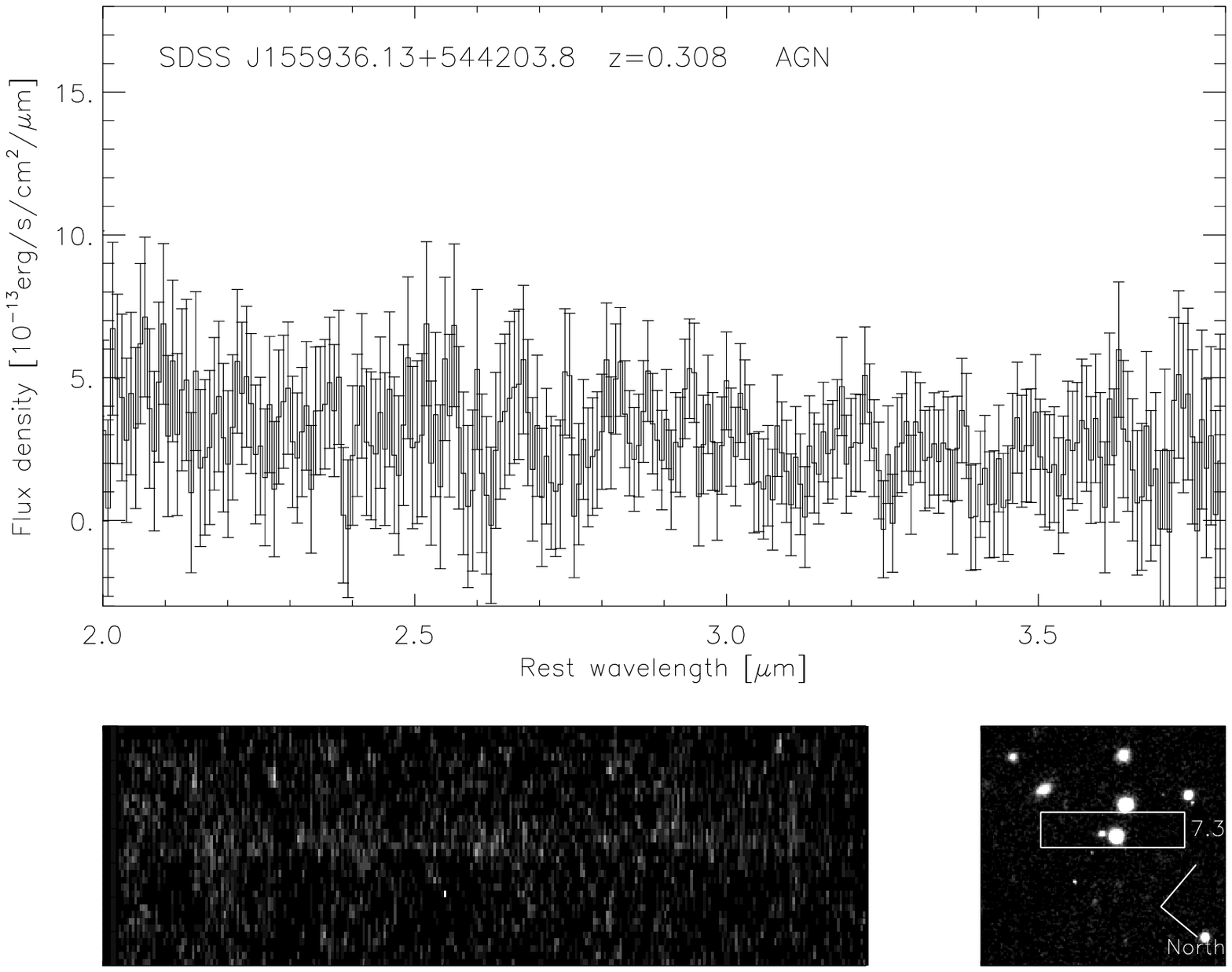}
\includegraphics[angle=90, scale=0.35]{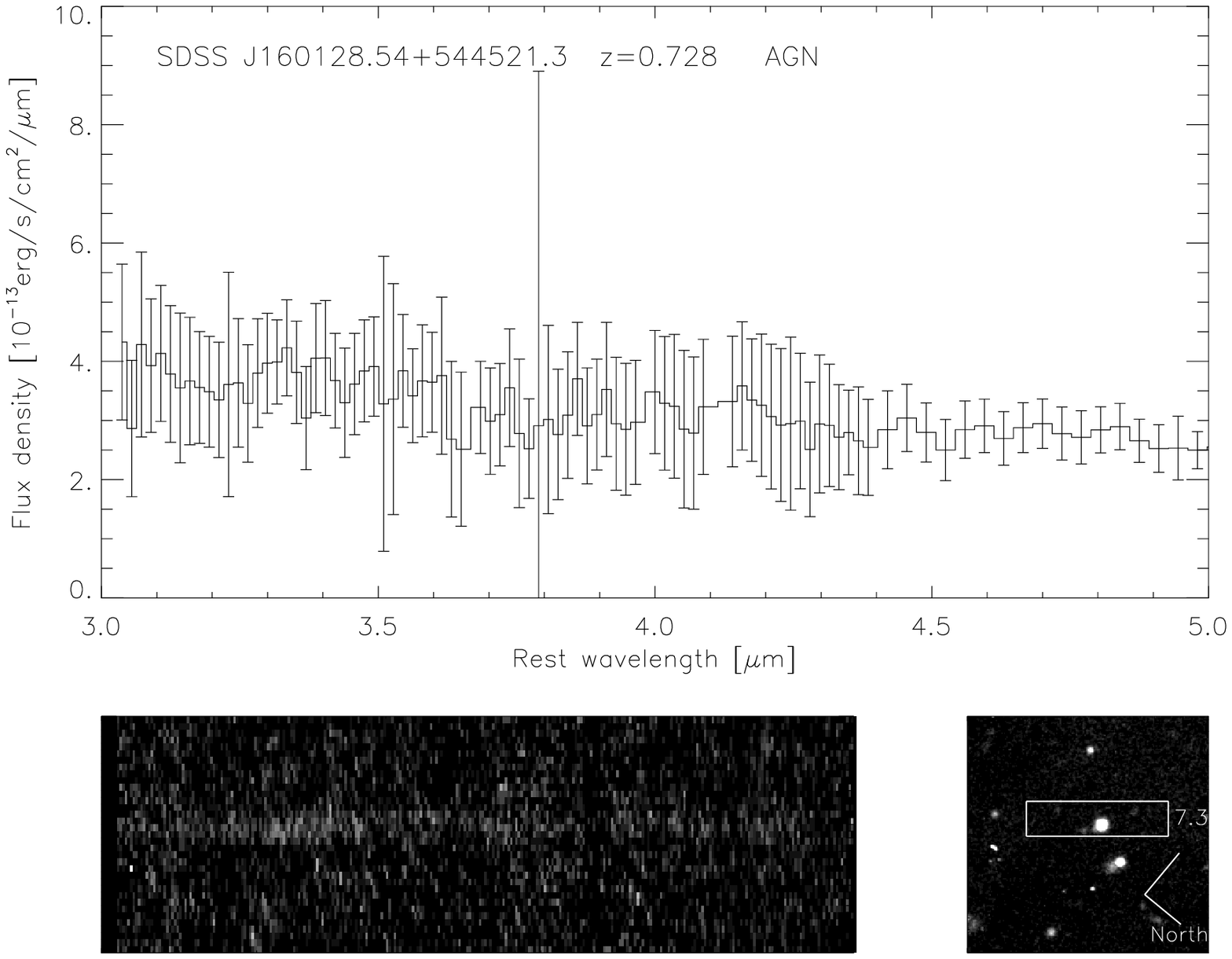}
\includegraphics[angle=90, scale=0.35]{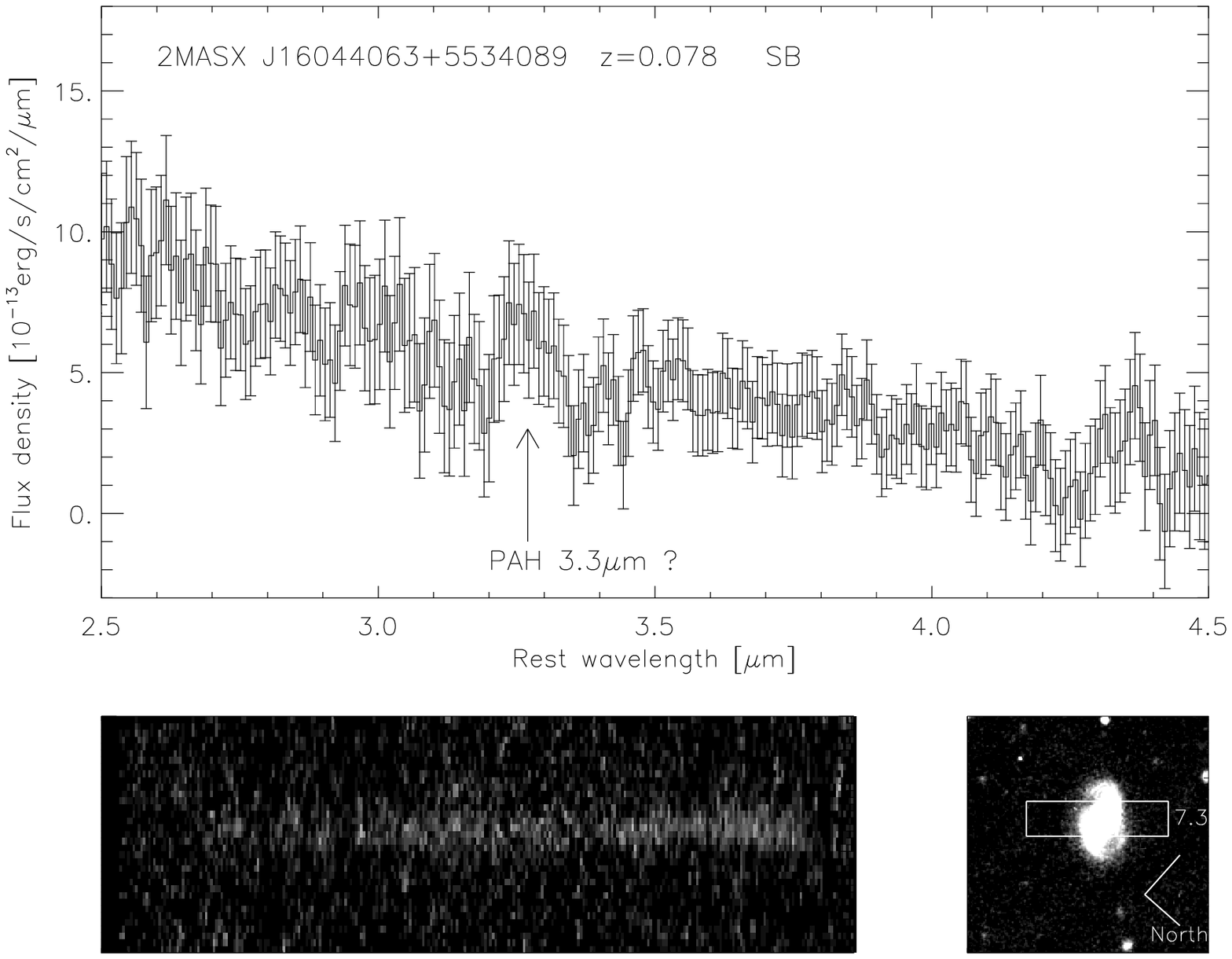}
\includegraphics[angle=90, scale=0.35]{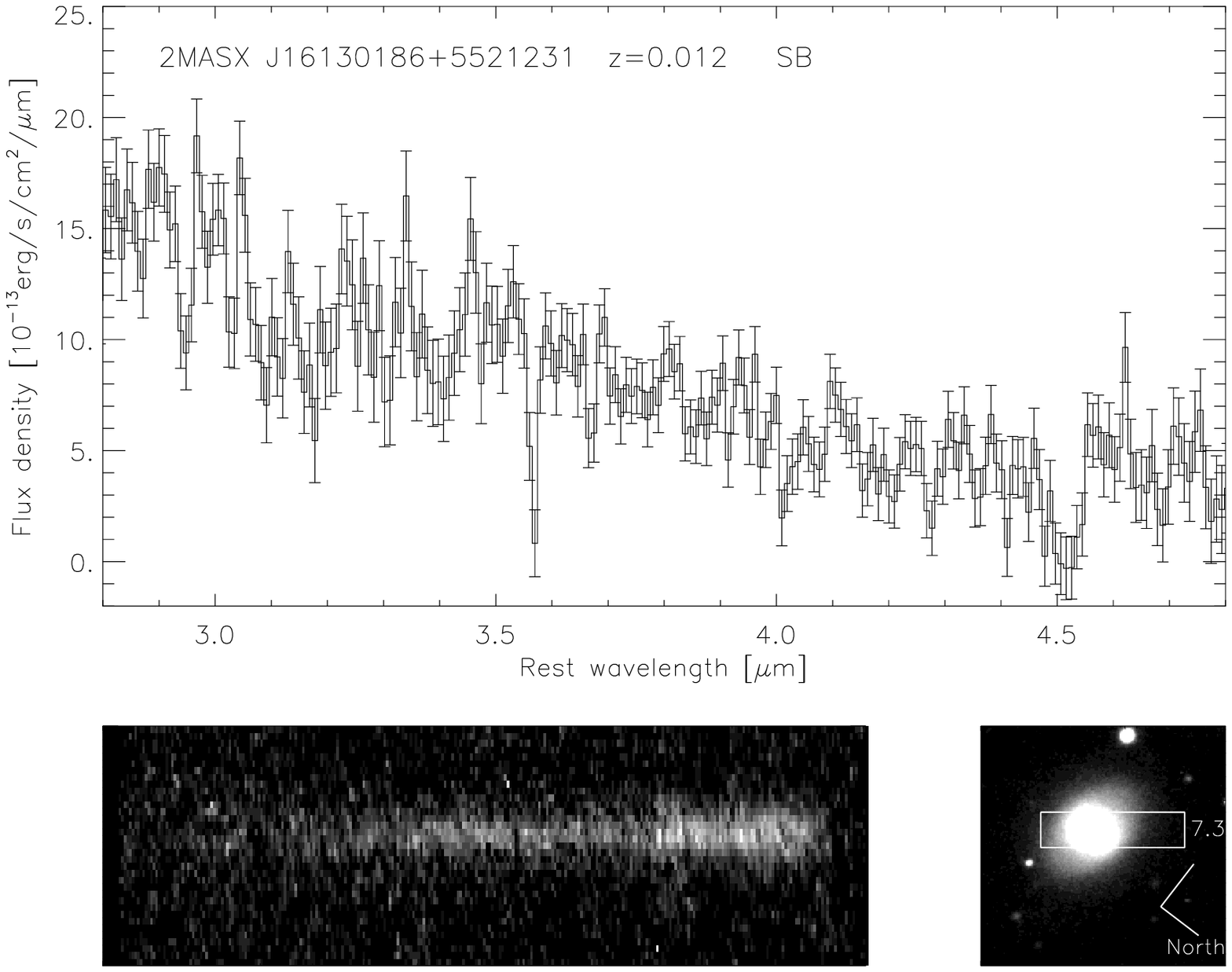}
\includegraphics[angle=90, scale=0.35]{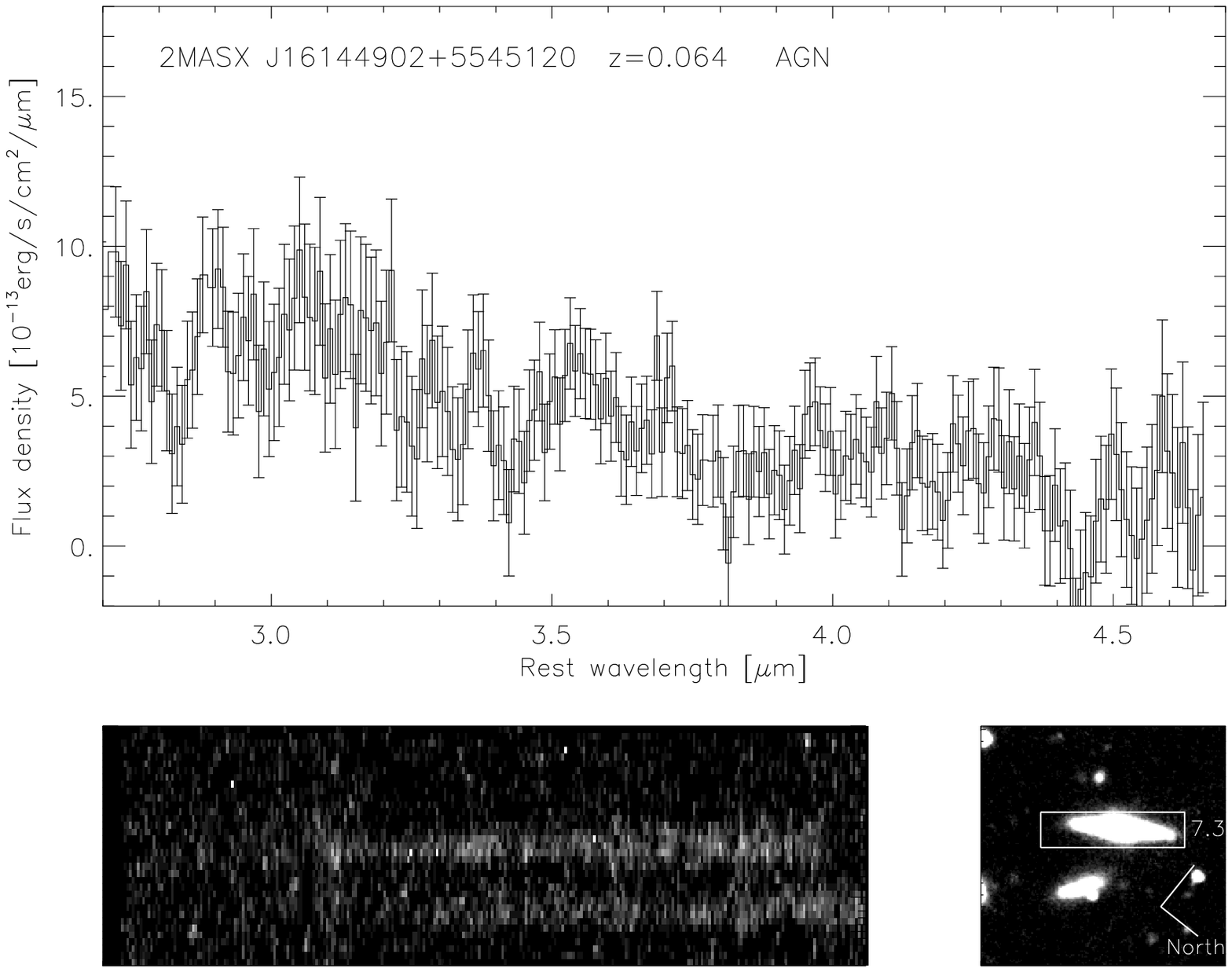}
\includegraphics[angle=90, scale=0.35]{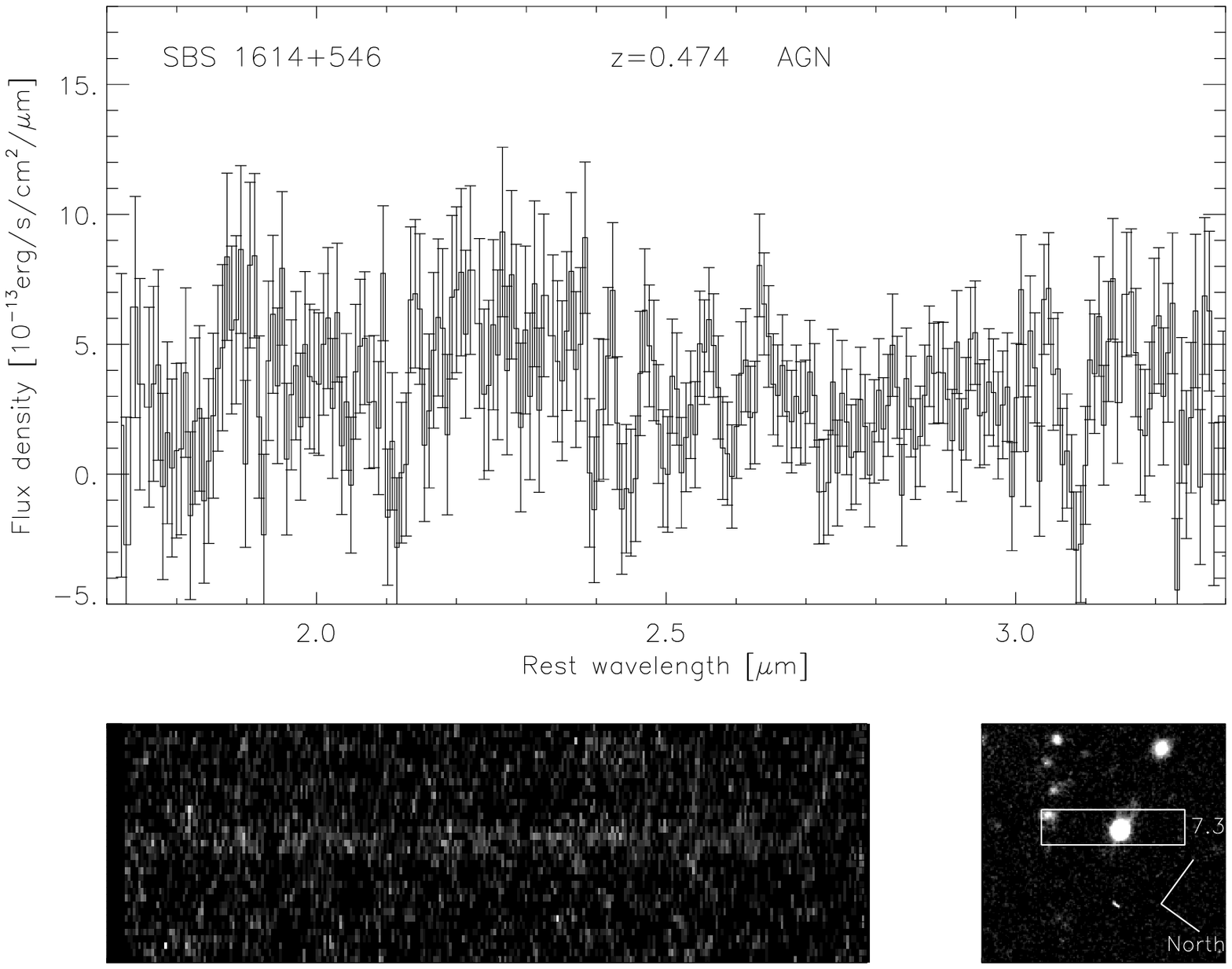}
\includegraphics[angle=90, scale=0.35]{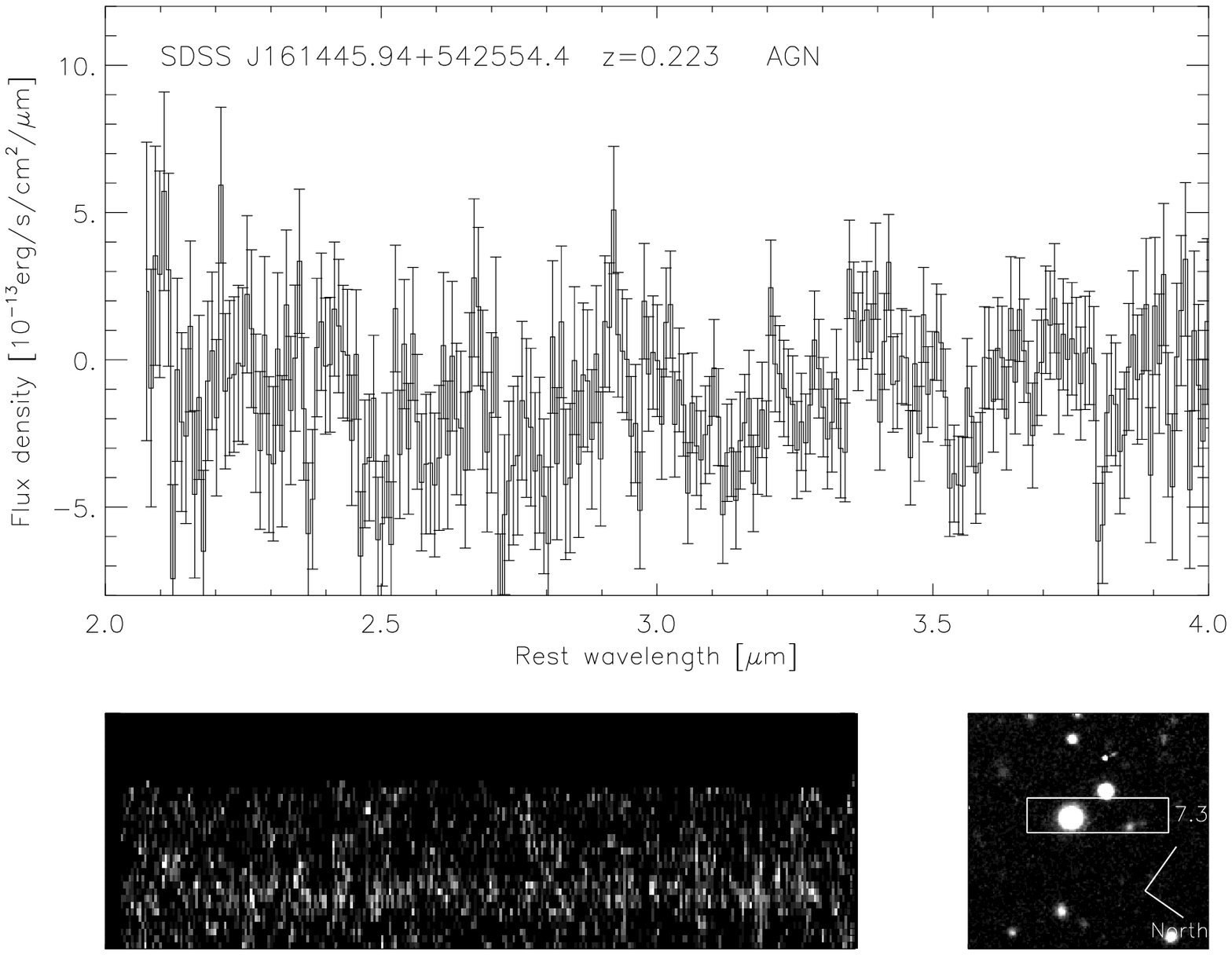}
\caption{Continued.
\label{fig2}}
\end{center}
\end{figure}

\begin{figure}
\figurenum{2}
\begin{center}
\includegraphics[angle=90, scale=0.35]{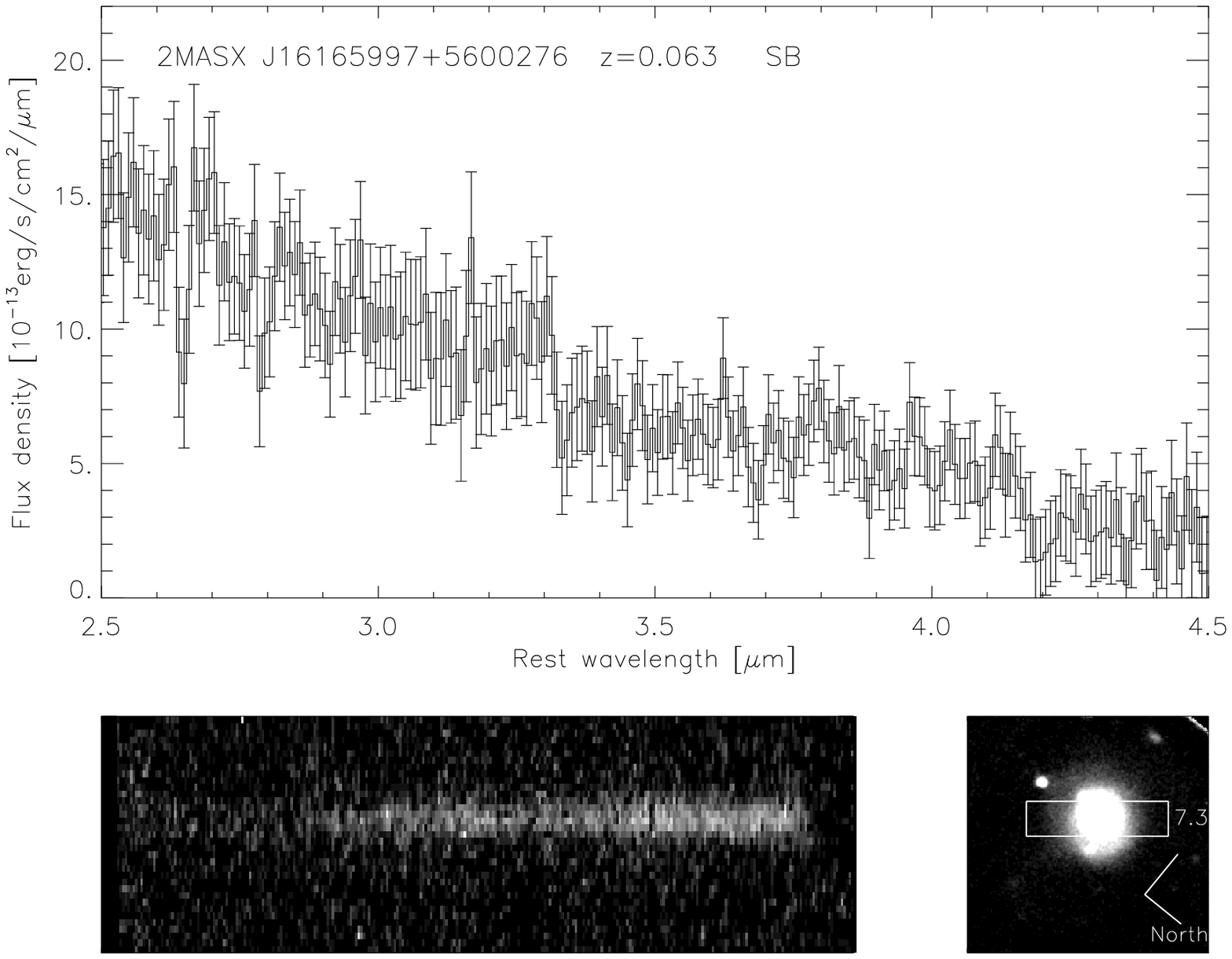}
\includegraphics[angle=90, scale=0.35]{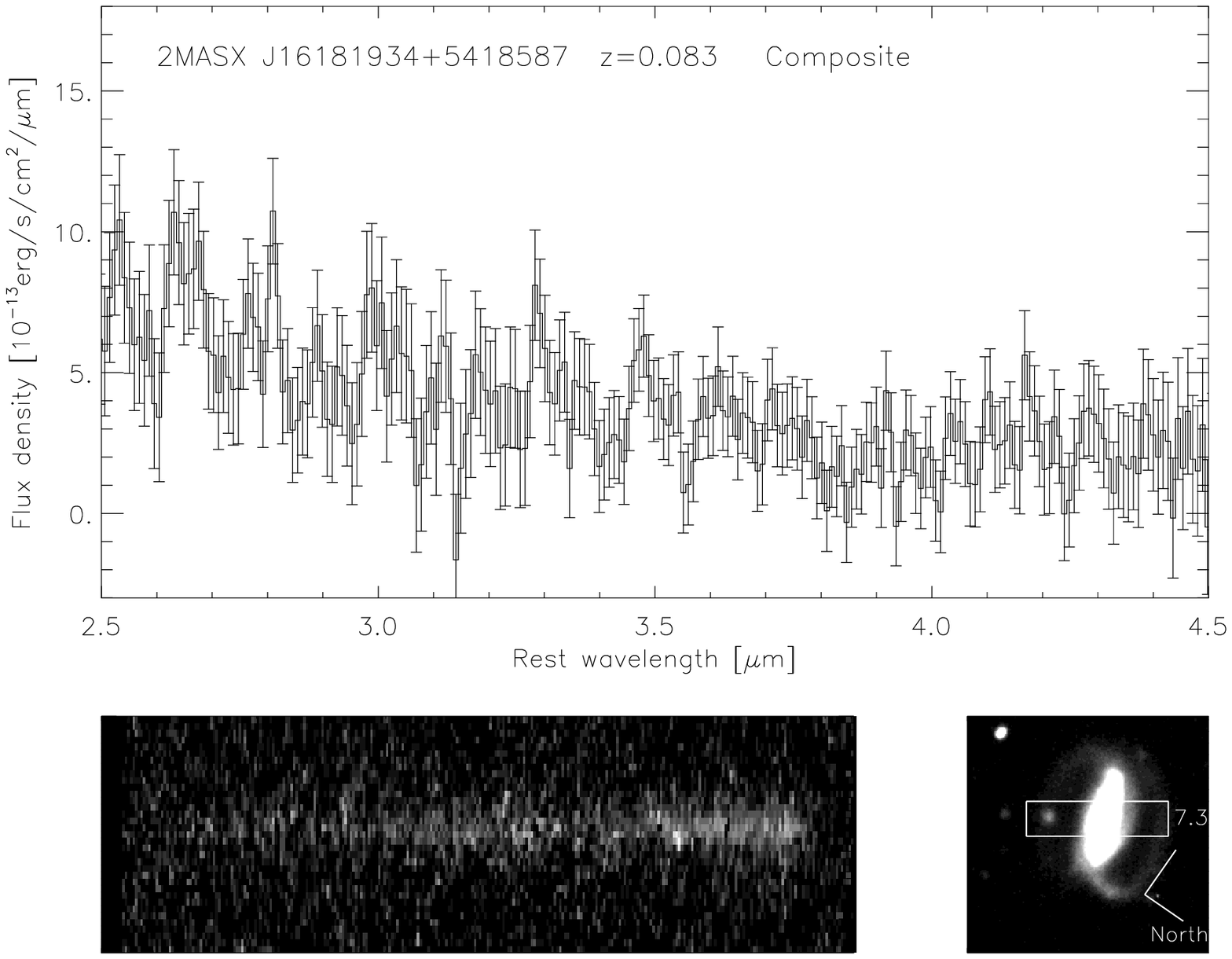}
\includegraphics[angle=90, scale=0.35]{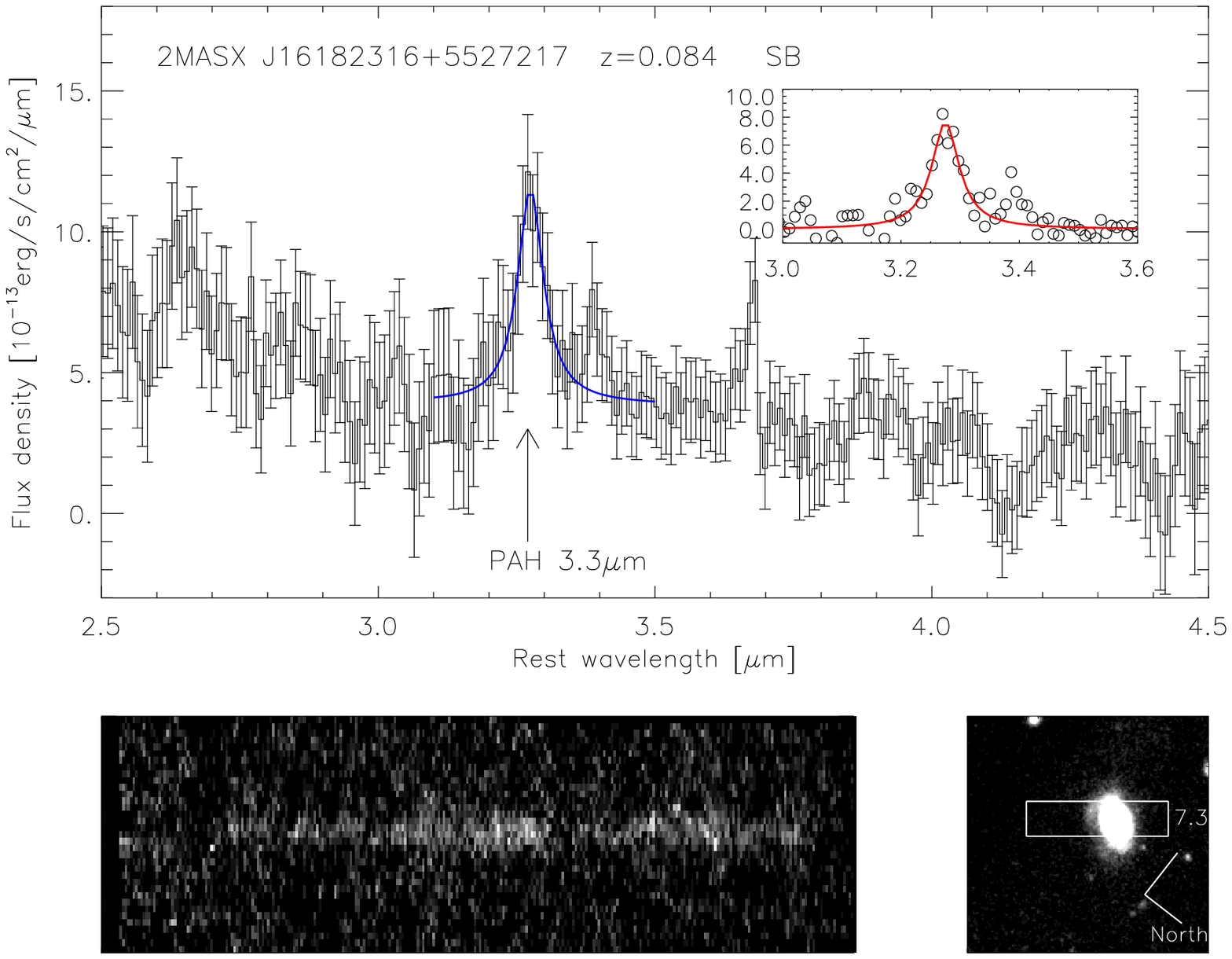}
\includegraphics[angle=90, scale=0.35]{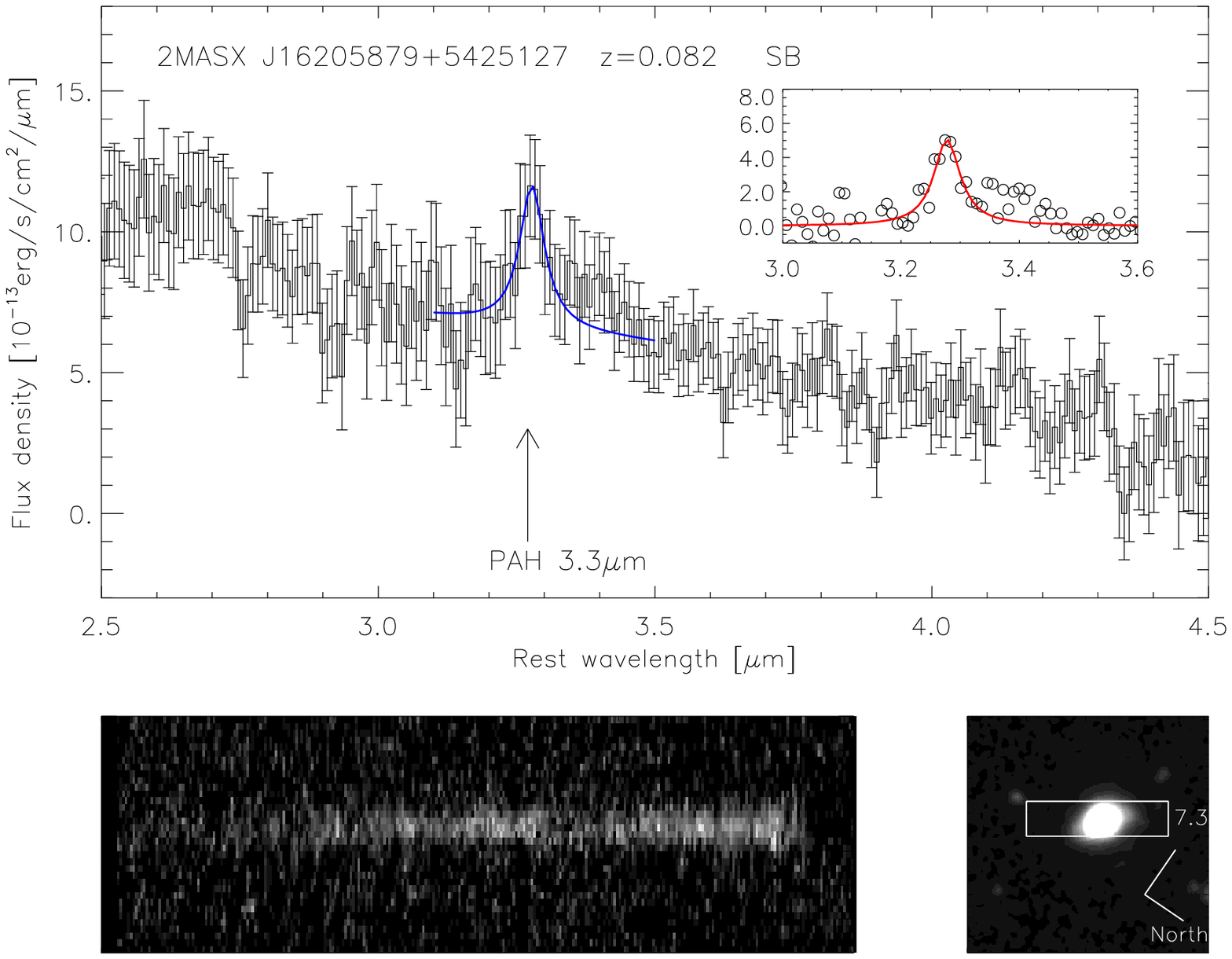}
\caption{The spectra of AMUSES based on IRC and IRS spectra along with ancillary images. All spectra are de-redshifted based on their redshifts, therefore x-axis is rest-frame wavelength. For two galaxies at high redshifts, FBQS J02191605-0557269 and SDSS J160128.54+544521.3, the spectra presented are based on IRS spectra. For the rest of the sample, the spectra are based on IRC spectra. Each panel includes the name of a target galaxy and its SED class as well. For galaxies detected with the 3.3 $\mu m$ PAH emission feature, we mark the locations of the feature by upward arrows and also show fitting results. Fitting results are shown with blue lines and insets. Overplotted blue lines are sums of continuum fits and fitted Drude profiles. Insets show the 3.3 $\mu m$ PAH emission feature after the fitted continuum subtracted from the spectrum. Then overplotted red lines within insets represent the fitted Drude profiles.
Lower left panels show the 2D spectra of target galaxies produced by IRC data reduction pipeline. The 2D spectra have the wavelength range of 2 - 5 $\micron$ which does not match the wavelength ranges of the 1D spectra presented. For these 2D spectra, the shorter wavelength end is on the right side opposite to the 1D spectra. Lower right panels show ancillary images of either the R-, or $Spitzer$ IRAC 3.6 $\micron$ band images. Six galaxies with $Spitzer$ IRAC 3.6 $\micron$ band images are FBQS J0216-0444, 2MASX J02165778-0324592, 2MASX J02191605-0557269, 2MASX J02193906-0511336, 2MASX J02195305-0518236, and 2MASX J16182316+5527217. Other than these six galaxies, all galaxies are with R-band images. 
These ancillary images are aligned to the 2D spectrum images to have the same spatial direction. Within the images, the extraction regions are drawn to the scale.
\label{fig2}}
\end{center}
\end{figure}

\begin{figure}
\begin{center}
\figurenum{3}
\includegraphics[angle=90, scale=0.7]{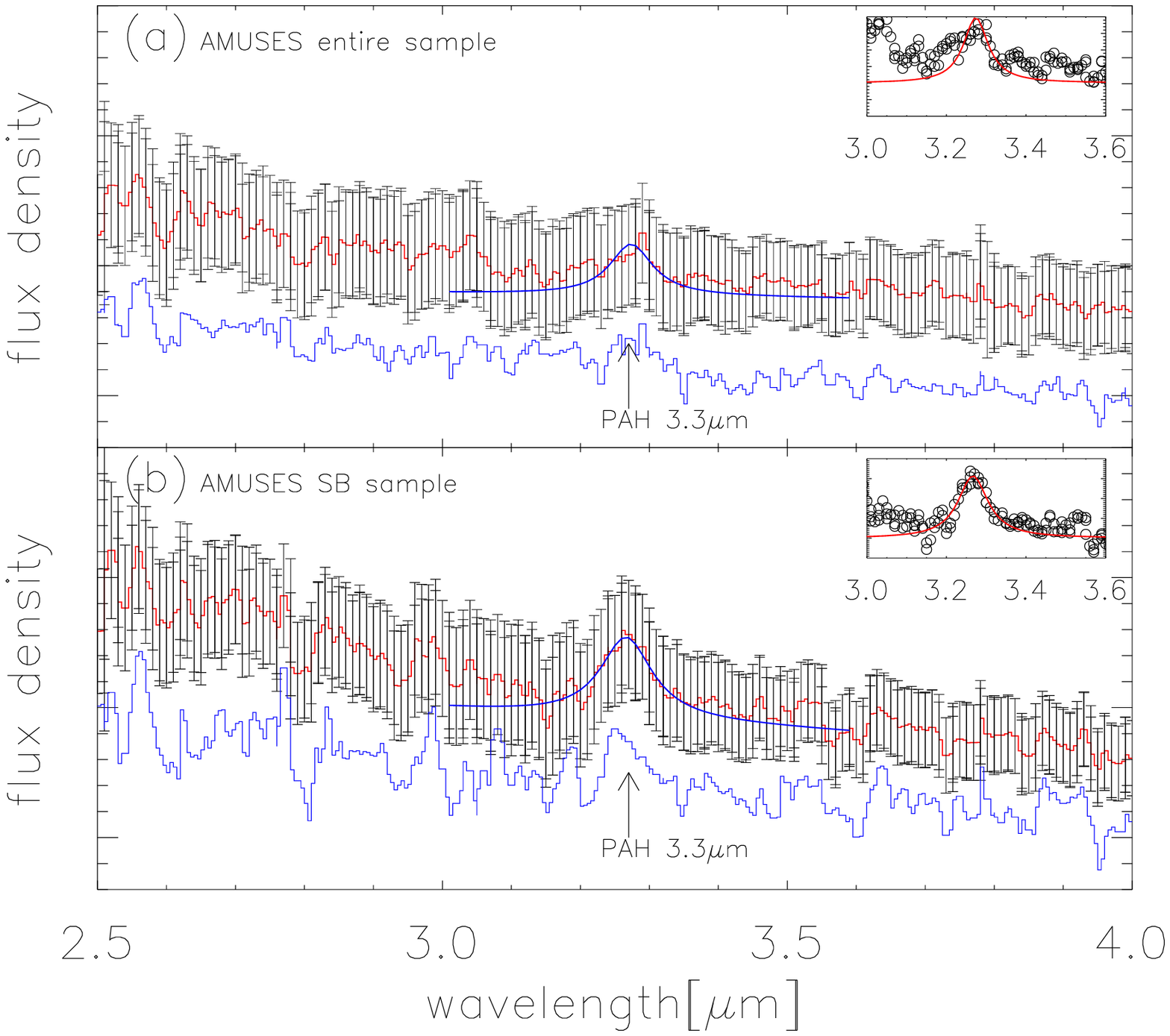}
\caption{The stacked spectra of subsamples. (a)  The spectrum stacked with the entire sample of 20 galaxies. (b) The spectrum spectrum stacked with the seven starburst SED galaxies. The normalized stacked spectra are overplotted for the entire sample and the starburst SED subsample respectively. These normalized stacked spectra are the mean of normalized individual spectra which are normalized by continuum flux within the wavelength range of 3.0 $\sim$ 3.6 $\micron$ in the rest frame. We adjust the flux levels of these normalized stacked spectra in order to avoid overlap between the spectra. We also show the fitting results for the stacked spectra. Overplotted blue and redlines, and the inset are the same with Figure \ref{fig2}.
\label{fig3}}
\end{center}
\end{figure}

\begin{figure}
\begin{center}
\figurenum{3}
\includegraphics[angle=90, scale=0.7]{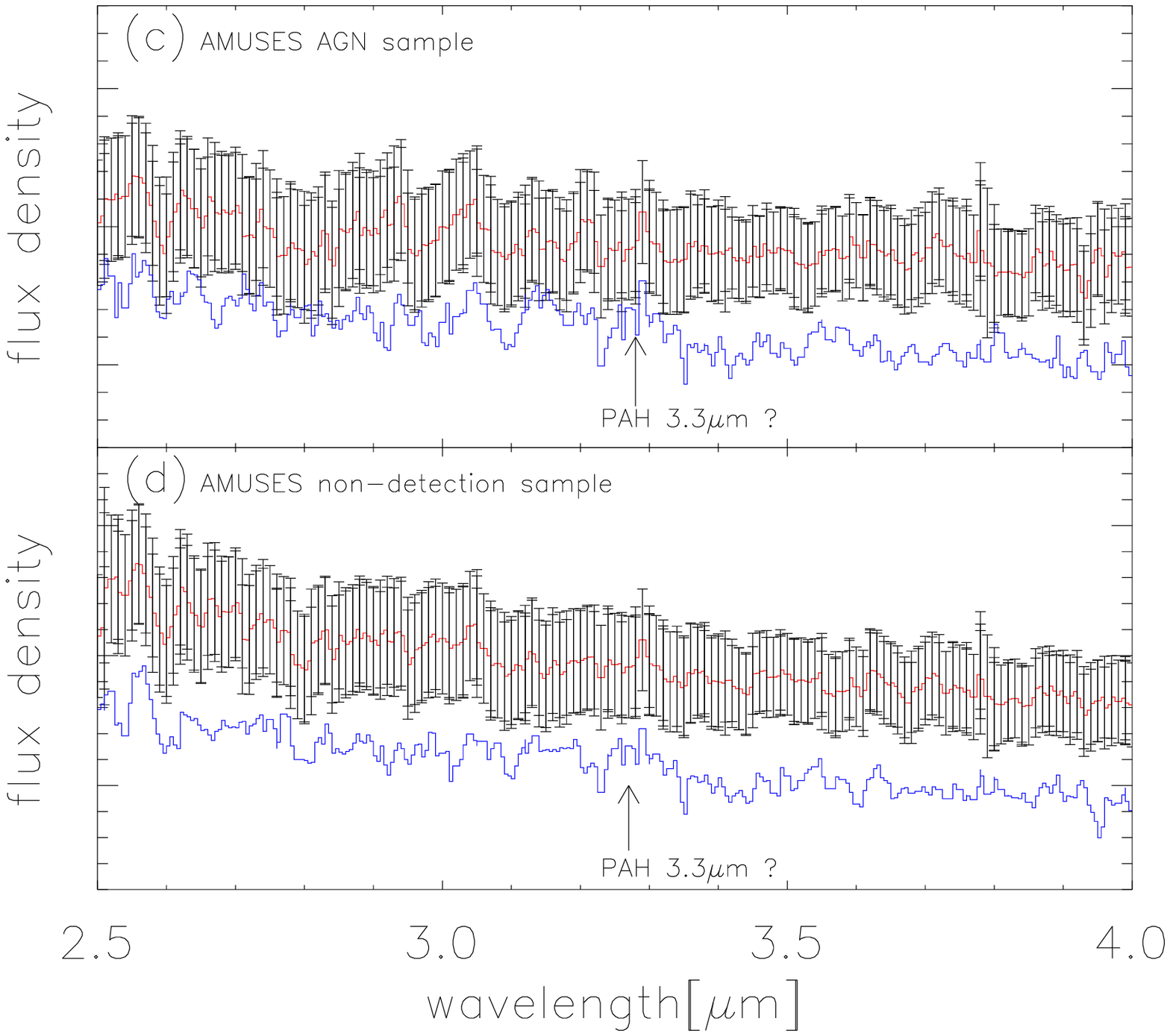}
\caption{The stacked spectra of the subsamples. (c) The spectrum stacked with the 12 AGN SED galaxies. (d) The spectrum stacked with the 15 non-detection galaxies. Overplotted blue spectra represent the normalized stacked spectra.
}
\end{center}
\end{figure}

\begin{figure}
\figurenum{4}
\begin{center}
\includegraphics[scale=0.6]{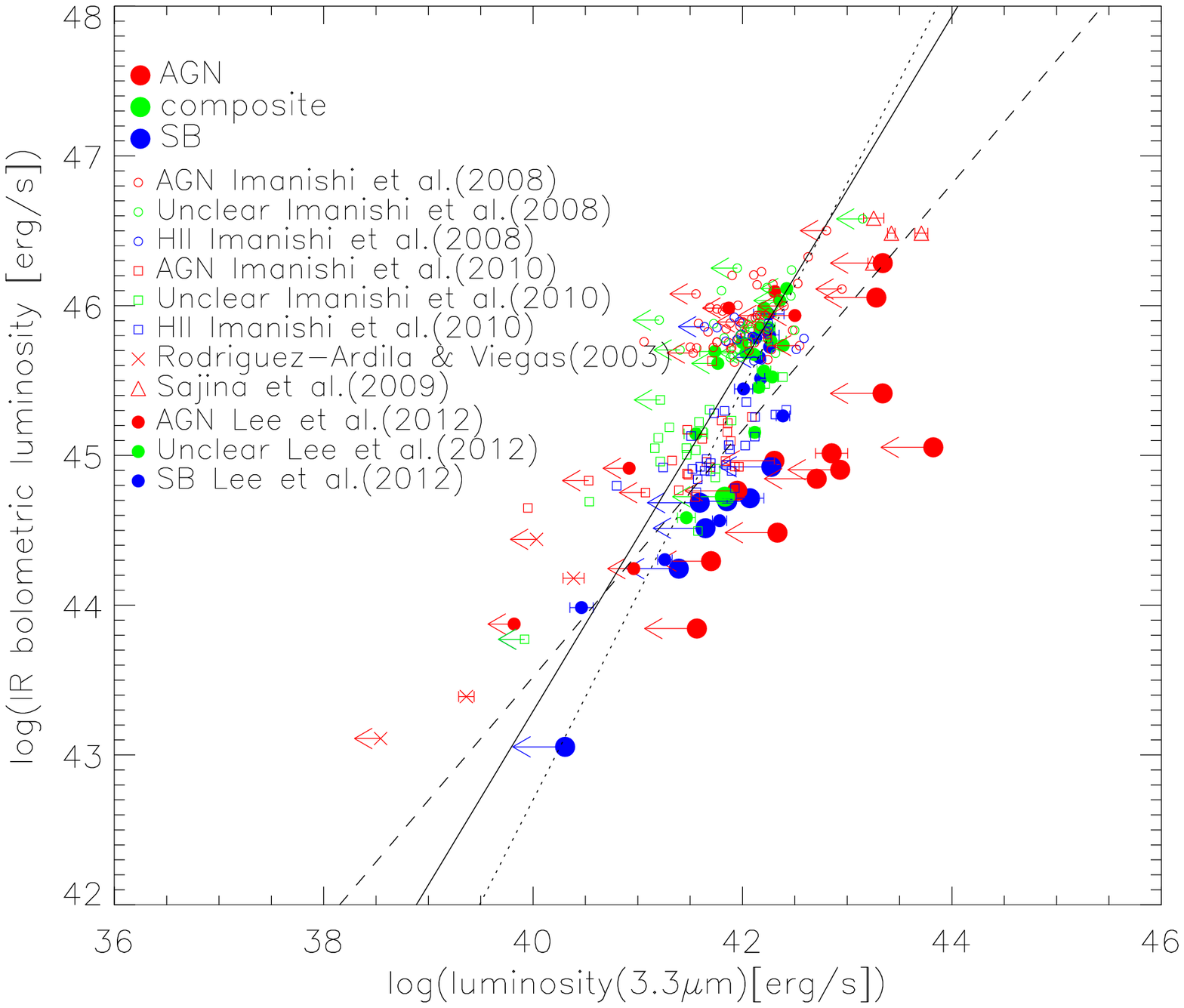}
\caption{Correlation between the luminosity of the 3.3 $\mu m$ PAH line and infrared luminosity. Bigger filled circles represent our sample. Blue circles represent galaxies with starburst SEDs while red circles represent galaxies with AGN SEDs. A green circle represent a galaxy with composite SED. The symbols without error bars, rather with leftward arrows represent upper limits. We describe how to decide these upper limits in \ref{line}. Overplotted are the data from \citet{Im08}, \citet{Sj09}, \citet{RAV03}, and \citet{Im10}.  The subsamples of \citet{Im08} and \citet{Im10} are represented by small empty circles and small empty squares, respectively. Their subsamples with AGN SEDs are represented by red symbols and their subsamples with HII-like SEDs are represented by blue symbols. Green symbols represent their subsamples with composite SEDs, or of uncertain SEDs. For the samples of \citet{Im08} and \citet{Im10}, data points with leftward arrows are upper limits as well. Small red crosses represent the sample of \citet{RAV03}, while small red empty triangles represent the high redshift sample of \citet{Sj09}.
The black solid line shows the linear fit for the 3.3 $\micron$ PAH emission detection from the combined sample of AMUSES and literature. The fit gives the the correlation between L$_{PAH 3.3 \micron}$ and $L_{\mathrm{IR}}$ as log ($L_{\mathrm{IR}}$) = (1.16 $\pm 0.30) \times$ log ($L_{\mathrm{PAH3.3}}$) - (3.11 $\pm$ 0.34). Then, the dotted line shows the linear fit to the data points with SB SEDs. The fit gives the correlation as log ($L_{\mathrm{IR}}$) = (1.37 $\pm 0.17) \times$ log ($L_{\mathrm{PAH3.3}}$) - (12.18 $\pm$ 0.75). Also presented is the fit to the non-ULIRG sample with SB SEDs. This fit gives the correlation as log ($L_{\mathrm{IR}}$) = (0.82 $\pm 0.13) \times$ log ($L_{\mathrm{PAH3.3}}$) + (10.58 $\pm$ 0.86). It is shown as the dashed line.
\label{fig4}
}
\end{center}
\end{figure}

\begin{figure}
\figurenum{5}
\begin{center}
\includegraphics[scale=0.6]{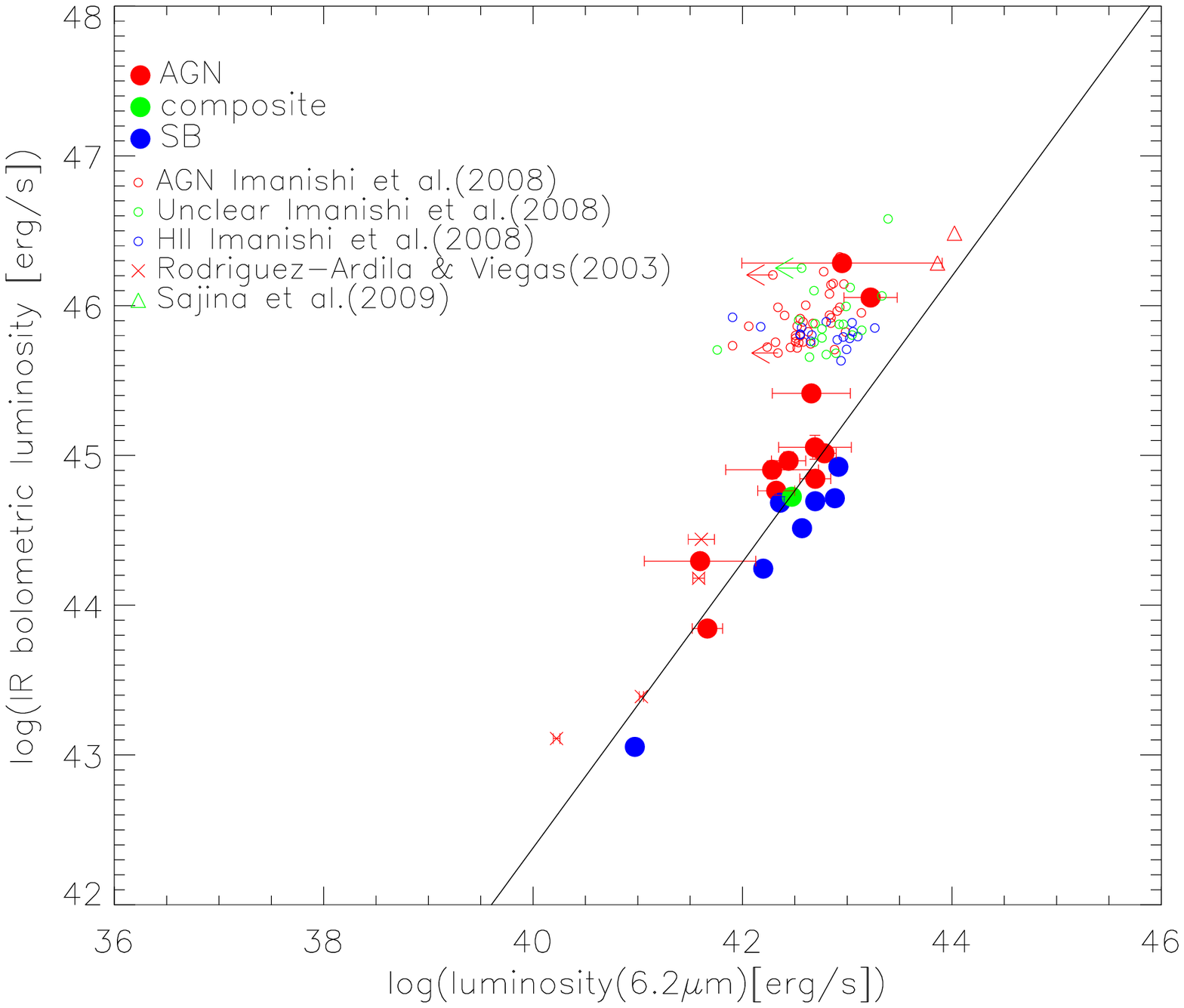} 
\caption{Correlation between luminosities of the 6.2 $\mu m$ PAH emission feature and infrared luminosity . The symbols are same with those of Figure \ref{fig4}. Again, the solid line is the linear fit which gives the correlation between these luminosities for the AMUSES sample.  It gives log ($L_{\mathrm{IR}}$) = (0.95 $\pm$ 0.03) $\times$ log ($L_{\mathrm{PAH6.2}}$) + (4.21 $\pm$ 0.19), where $L_{\mathrm{IR}}$ and $L_{\mathrm{PAH6.2}}$ are in the unit of erg sec$^{-1}$.
\label{fig6}
}
\end{center}
\end{figure}

\begin{figure}
\figurenum{6}
\begin{center}
\includegraphics[scale=0.6]{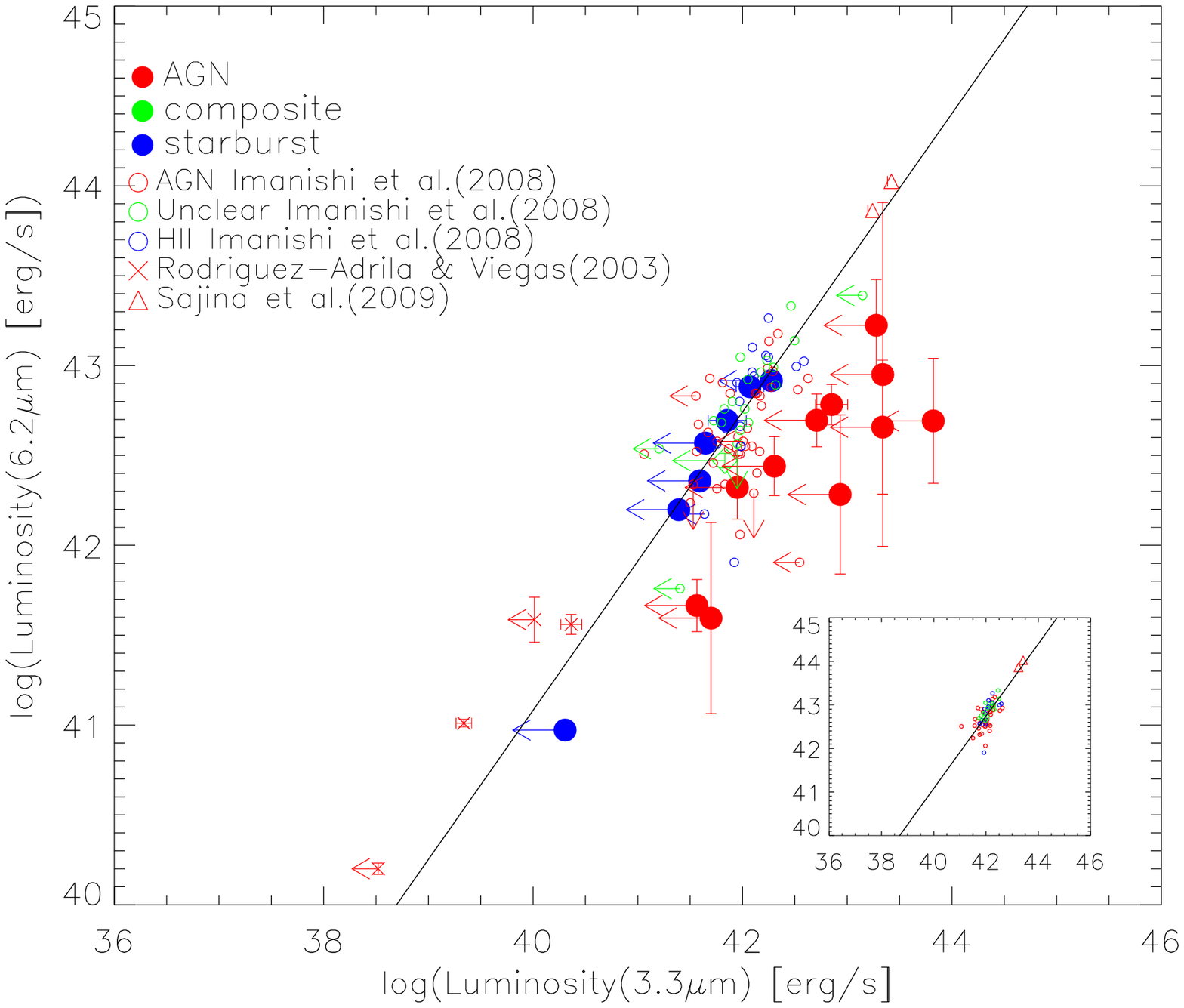}
\caption{Correlation between luminosities of the 3.3 $\mu m$ PAH emission feature and the 6.2 $\mu m$ PAH emission feature. The symbols are same with those of Figure \ref{fig4}. Again, the black line is the linear fit which gives the correlation between these two PAH luminosities for all the data points excluding the upper limits.  It gives log (L$_{PAH 6.2}$) = (0.83 $\pm$ 0.06) $\times$ log ($L_{\mathrm{PAH3.3}}$) + (7.88 $\pm$ 0.41), where $L_{\mathrm{PAH3.3}}$ and $L_{\mathrm{PAH6.2}}$ are in the unit of erg sec$^{-1}$ The small inset shows 66 sources with L$_{IR} >$ 10$^{12}$ L$_{\odot}$. The fit within the inset is same with the one given for the entire sample.
\label{fig5}
}
\end{center}
\end{figure}



\end{document}